%% file: paper.tex
\documentclass{egpubl}
\usepackage{eg2021}
 
\ConferencePaper        
\usepackage[T1]{fontenc}
\usepackage{dfadobe}  

\usepackage{cite}  
\BibtexOrBiblatex
\electronicVersion
\PrintedOrElectronic
\ifpdf \usepackage[pdftex]{graphicx} \pdfcompresslevel=9
\else \usepackage[dvips]{graphicx} \fi

\usepackage{egweblnk} 

\usepackage{balance}
\graphicspath{{figures/}{pictures/}{images/}{./}} 
\newcommand{\engine}{Continuum}
\newcommand{\game}{Across Dimensions}

\title[Higher Dimensional Graphics]%
      {Higher Dimensional Graphics: \\Conceiving Worlds in Four Spatial Dimensions and Beyond}

\author[M. Cavallo]
{\parbox{\textwidth}{\centering M. Cavallo \orcid{0000-0003-1506-4536}} \\
{\parbox{\textwidth}{\centering IBM Research, Yorktown Heights, NY}}
}

\begin{document}


\maketitle
\begin{abstract}

\input{abstract}

\end{abstract}  

\input{introduction}
\input{related}

\input{method_1}

\input{method_2}

\input{usecase}


\input{conclusion}

\bibliographystyle{eg-alpha-doi} 
\balance
\bibliography{paper}       



\end{document}

%% file: abstract.tex
While the interpretation of high-dimensional datasets has become a necessity in most industries, 
the \textit{spatial} visualization of higher-dimensional geometry has mostly remained a niche research topic for mathematicians and physicists.
Intermittent contributions to this field date back more than a century, and have had a non-negligible influence on contemporary art and philosophy. However, most contributions have focused on the understanding of specific mathematical shapes, with few concrete applications. 
In this work, we attempt to revive the community's interest in visualizing higher dimensional geometry by shifting the focus from the visualization of abstract shapes to the design of a broader hyper-universe concept, wherein 3D and 4D objects can coexist and interact with each other.
Specifically, we discuss the content definition, authoring patterns, and technical implementations associated with the process of extending standard 3D applications as to support 4D mechanics. 
We operationalize our ideas through the introduction of a new hybrid 3D/4D videogame called \textit{\game}, which we developed in Unity3D through the integration of our own 4D plugin.

%% file: introduction.tex
\section{Introduction}

Conceiving of and visualizing a reality characterized by more dimensions than a human can perceive represents a long-lasting problem that is both intellectually and computationally difficult to solve.
Mathematicians have generalized tridimensional concepts, formulating the existence of higher dimensional geometries such as knotted spheres \cite{banchoff2002computer}, while physicists have theorized phenomena occuring in higher dimensional spaces (e.g. superstring and quantum theories) \cite{mcmullen2008visual}. Literature, philosophy, and art have also contributed to the notion of extra spatial dimensions \cite{henderson2013fourth}, offering such conceptions as the neoplatonic concept of ``cosmic consciousness'' and Dante's hyperspheric cosmos in ``La Divina Commedia'' \cite{egginton1999dante}, as well as theological literature suggesting the existence of divine entities in higher dimensions \cite{willink1893world}.
Projection into two- and three-dimensional spaces is the standard technique for enabling humans to experience higher dimensional objects, but comes at the cost of introducing ambiguity and losing geometrical information such as symmetry and curvature \cite{cavallo2018visual}.
Over the years, computer graphics and visualization research has attempted to enhance intuitive user experiences of these abstract geometric shapes through the introduction of visual cues (e.g. depth perception, colors, shadows), new interactive methods of manipulating 4D objects, and novel display methods \cite{hanson1993interactive,mcmullen2008visual,yan2015new}.

In this work, rather than trying to improve upon existing visualization or projection techniques for specific geometry, we aim to present a comprehensive approach of how standard 3D graphics pipelines can be extended as to conceive and interface higher dimensional, and more specifically, four-dimensional worlds. By ``dimensions'' we refer to the independent directions into which a geometrical space extends---not including \textit{time}.
We start by introducing a simple 4D rendering engine called \textit{\engine} (implemented as a Unity3D plugin), and we observe how basic concepts such as movement and interaction performed in higher dimensions manifest for users viewing projections of these apparently ineffable worlds.
We then draw on the laws of physics to lay out possible ways of structuring a higher dimensional universe, and identify which content generation and authoring methods could be used to populate such a universe.
We qualitatively evaluate our work through a new videogame called \textit{\game}, showcasing how 4D and 3D elements can be seamlessly combined in a concrete application.
Finally, we conclude by discussing the challenges and potential benefits of applying higher dimensional concepts to existing 3D graphics applications.

%% file: related.tex
\section{Related Work}
In the first part of this section we focus on the formulation of concepts of \textit{higher dimensions} in the twentieth century, noting how this field lies at the intersection of science, art, and philosophy. Then we review more recent computer graphics and data visualization attempts at handling 4D geometry.

\subsection{A Brief History of Higher Dimensions}
During the last two decades of the nineteenth century, scientific innovations such as Röntgen's  discovery of the X-ray, Rutherford's and Curie's work on radioactivity, Thomson's identification of the electron, and Hertz's studies on electromagnetic waves profoundly altered conceptions of matter and space, radically expanding our understanding of the physical world and acknowledging the limits of human perception.
Theories on the existence of a suprasensible reality
began to thrive \cite{henderson2013fourth,zollner1882transcendental,stewart1878unseen}, and our world was first described as a mere boundary of an unperceivable higher-dimensional reality \cite{stewart1878unseen}.
The first major popularization of a spatial fourth dimension can be found in Abbot's 1884 novel \textit{Flatland} \cite{abbot1884flatland}, wherein the author describes how inhabitants of a fictional two-dimensional world might refuse to accept the possibility of higher dimensions.
Distortion, denial of the three-dimensional perspective, multiple vanishing points, and high dimensional space continuity soon became recurring elements in the ineffable realities depicted by artists such as Picasso and Boccioni \cite{henderson2013fourth}.
Einstein's 1905 Special Theory of Relativity \cite{miller1998albert} and Minkowski's 1908 formulation of the \textit{space-time continuum} \cite{petkov2010space}, however, led to a gradual transformation of the concept of the fourth dimension from \textit{space} to \textit{time}.
In the geometrical interpretation known as the \textit{block universe} \cite{jeans1931mysterious}, space and time are geometrically welded together in a four-dimensional volume, and human life simply progresses through 3D cross-sections of this higher dimensional structure.
Playful speculation on four-dimensional geometry continued to appear in the work of artists such as Davis, Dalì and Duchamp \cite{banchoff2014salvador}, and evolved into the concepts of \textit{hyperspace} and \textit{cyberspace} in cinematography and science-fiction \cite{heinlein1941and,nahin2001time,youngblood2020expanded, kaku2016hyperspace, wertheim2000pearly,benedickt1991cyberspace}.
While a fascinating research direction, the interpretation of time as a geometrical dimension in its own right will not be explicitly covered in this work.

It was only towards the end of the twentieth century that the concept of a fourth dimension was reinterpreted as a suprasensible dimension of \textit{space}, within which our 3D world would exist merely as a section or boundary. The first world conference on four-dimensional graphics (or ``hypergraphics'') was held in 1984 \cite{banchoff1990flatland}, on the occasion of the centennial anniversary of the publication of Flatland.
During the same year, \textit{superstring theory} emerged in physics as a possible ``theory of everything'', and attempted to achieve a grander unification scheme by coupling gravitation and electromagnetism, supporting the existence of up to eleven compactified dimensions of space \cite{paul1984eleventh}.
The advent of digital media and personal computers, with their increased graphics and visualization capabilities, were among the main contributors to this reignited interest in higher spatial dimensions. Michael Noll \cite{noll1967computer} had already demonstrated in 1967 how a computer can easily construct higher dimensional content, while Arnold \cite{arnold1972proposal} had proposed initial studies of how the human brain might retain 4D information \cite{von2003cybernetics}.
The field of four-dimensional computer graphics flourished between the 1980s and the 1990s through the work of authors such as Banchoff and Strauss \cite{banchoff1978hypercube, banchoff1990beyond, brisson1987hypergraphics}, and was accompanied by emerging art, cinematic and architectural endeavors to depict an invisible, ``higher'' reality \cite{rucker2012geometry,rucker2014fourth,robbin1992fourfield,novak1998transarchitectures}.
A decade later, superstring theory would be accompanied and later superseded by M-theory \cite{duff1996m}, according to which extra-dimensions are compactified in curled-up membranes \cite{passages2006unravelling}. 


\subsection{Recent Attempts at 4D Geometry Visualization}
All methods of representing 4D content to the human eye, either on a flat screen or in virtual environments, require reducing the dimensionality of the data. To compensate for the loss of visual information caused by this process, researchers have developed over time a set of dedicated visualization techniques.
Early work on higher dimensional rendering focused on the use of color, texture, shadows and lighting to indicate 4D depth and create intuitively informative shaded images of mathematical 4D objects \cite{hanson1992illuminating,hanson1993interactive,hausmann1994visualization}.
Grid planes and hypervolume removal techniques were used to visualize vector fields  \cite{hanson1992four,bajaj1998hypervolume}, whereas the visualization of silhouettes, auxiliary vectors, and surface normals was used to improve the  understanding of spatial relationships in hypersurfaces\cite{hoffmann1991some,banks1992interactive,miwa20174}.
Other experiments include attempts at ``voxelizing'' 4D geometry through hypercubes \cite{balsys2007point}, constructing isosurfaces through convex hulls \cite{bhaniramka2004isosurface}, adding volumetric halos to hyper-surfaces \cite{wang2013interactive}, and visualizing trajectories of high-dimensional dynamical systems through 3D parallel coordinates \cite{wegenkittl1997visualizing}.
While parallel research focused on \textit{time} as the fourth dimension, and attempted to visualize 3D movements or shape changes in a flat space-time continuum \cite{sakai2007four, ohori2017modeling, ohori2017visualising,ohori2013representing}, \textit{time} was also incorporated into the understanding of 4D geometry through real-time, interactive manipulation of 4D objects \cite{hanson1993interactive}.
This was progressively made possible thanks to the improvements in rendering performance and hardware optimization, and to the introduction of the first GPU-accelerated tetrahedron-based rendering pipelines \cite{chu2009gl4d}.

Based on the notion that an understanding of shapes benefits from a combination of sight, touch, and other senses \cite{zhang2014visualizing}, various multimodal interactive manipulation approaches have been proposed. Examples include the use of multitouch interactions \cite{yan2012multitouching,zhang2014visualizing}, tangible interfaces \cite{li2015and}, haptics and force-feedback \cite{zhang2007shadow}, standard and custom controllers \cite{hanson1999meshview,kageyama2016keyboard,sakai2011four}, and even user positional input \cite{aguilera2006virtual}. Particular attention has been paid to visual and sensory cues allowing one to visualize 4D collisions and to identify smooth, non-intersecting hypersurfaces (e.g. knotted surfaces embedded in 4D), whose 3D projections are degraded by self-intersections \cite{hanson2005multimodal,zhang2007shadow}. Visualization of often semi-transparent, ``ghost geometries'' has also been used to preview the projection effects of 4D manipulations \cite{li2015and,kageyama2016visualization}, together with the application of multiple simultaneous projections \cite{matsumoto2019polyvision}.
Several studies on spatial perception have also been performed to assess if and how the human brain could understand higher dimensions, and form mental images of such spaces \cite{miwa20154}. Ambinder et al. \cite{ambinder2009human} have shown evidence that people can learn to make spatial judgments on the length of, and angle between, line segments embedded in 4D space, while Ogmen et al. \cite{ogmen2019perception}  studied the inherent capabilities and limitations of brain plasticity in building perceptual and sensorimotor 4D representations. One interesting experiment involved assessing whether a human could find an exit from an n-dimensional labyrinth \cite{jamroz2009multidimensional}---representing a rare attempt at creating a multidimensional computer game.
Seeking to improve 4D perception through stereoscopic vision, researchers moved from using standard flat screens to the adoption of glass-free 3D displays \cite{sakai2011four}, immersive environments such as CAVE systems \cite{cross1994virtual,aguilera2006virtual}, and augmented or virtual reality devices \cite{li2015and,kageyama2016visualization,matsumoto2019polyvision}.
Recent work by M.T. Bosch has focused on physics applied to four-dimensional ``toys''  \cite{bosch2020n}, generalizing the rules of how objects bind together, slide, and roll to higher dimensions. The videogame Miegakure \cite{miegakure}, currently under development, leverages tetrahedral meshes and cross-section to provide a ``hide and reveal'' puzzle/adventure experience, and represents one of the possible use cases covered by our work.\\
Different from existing literature, this paper attempts to go beyond the visualization of isolated, abstract 4D objects, instead conceiving of a more general 4D universe, where multiple 4D and 3D entities can coexist and interact with each other in concrete use cases.
We also note how prior work generally relies on either cross-section (mainly used for interactive manipulation) or frustum projection (more common in illustrations or educational applications), with very few examples \cite{miwa2013interactive,miwa2013four} mentioning camera use and the control of vanishing points. Here we use both cross-section and frustum projection methods, and compare their usage in interactive applications where the \textit{user}---not the 4D model anymore---is the protagonist.

%% file: method_1.tex
\begin{figure}[t]
\centering
 \includegraphics[width=0.9\linewidth]{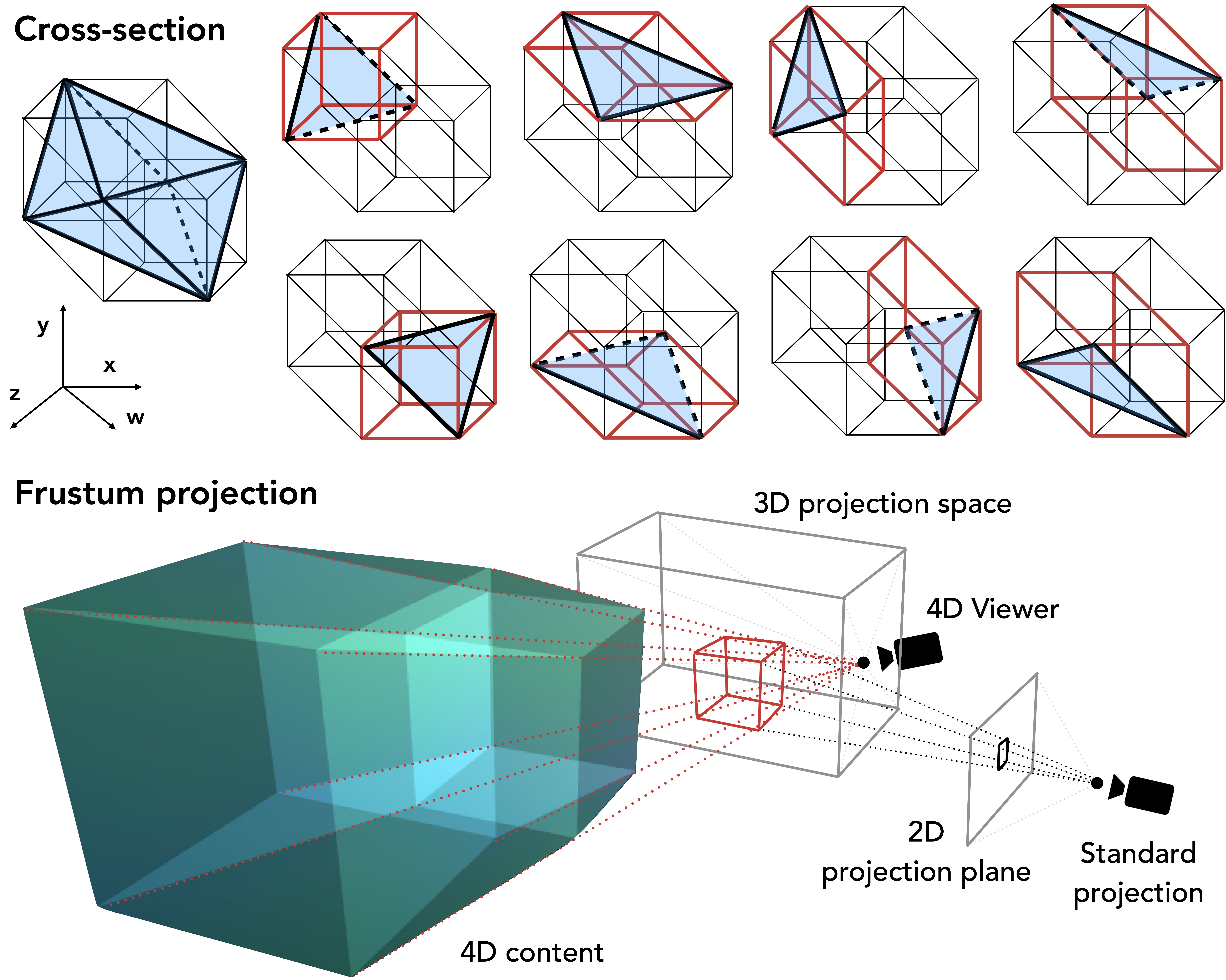}
   \vspace{-0.5em}
 \caption{\textbf{Projection methods}. Cross-section slices 4-dimensional objects based on their intersection with an arbitrary hyperplane. In this example, the cross-section of a rotated tesseract is constructed by merging the polyhedra generated by the intersection of its 4D geometry with a hyperplane. Frustum projection (or perspective) ensures objects further away from the viewer along $w$ appear smaller. We implement this through the use of a 4D camera which generalizes the standard frustum projection, and generates the 3D meshes that will be later rendered by the usual 3D camera.}
 \label{fig:projection}
 \vspace{-1em}
\end{figure}

\begin{figure}[t]
\centering
 \includegraphics[width=1\linewidth]{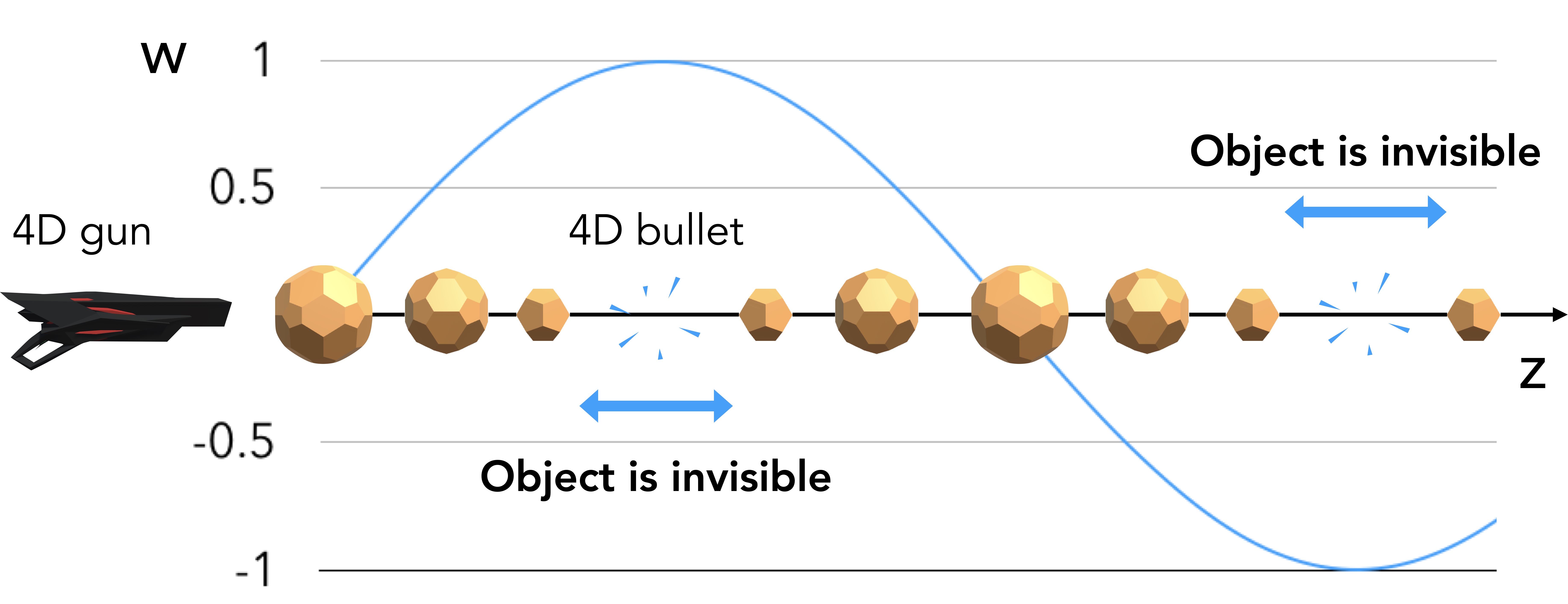}
   \vspace{-1.5em}
 \caption{\textbf{4D Translation}. When a 4D object moves along the fourth dimension, its geometry is differently sliced based on the current cross-section hyperplane, producing a visual effect similar to morphing. When there is no intersection with the hyperplane, the projection does not generate a mesh, making the 4D object ``disappear.'' If we imagine shooting a bullet with a hypothetical 4D gun, generating a sinusoidal movement along $w$, the perceived effect would be a bullet continuously entering and exiting the cross-section (i.e. appearing and disappearing).}
 \label{fig:bullet}
 \vspace{-1em}
\end{figure}

\section{\engine}

In this section we briefly introduce \engine, our 4D rendering engine, which has been used to perform the experiments and to render the figures shown in this paper. The development of \engine ~was not aimed at creating a fully featured, state-of-the-art, or high performance engine. Rather, it was conceived as very minimal, and focused on achieving our research goals. We implemented \engine ~as a Unity3D plugin, which allowed us to leverage the tools already offered by the popular 3D game engine and make \engine ~more accessible.
Below we briefly describe how our engine extends the standard 3D approach to geometry definition, projection, and rendering, emphasizing how these newly introduced 4D concepts are perceived by the viewer. We refer our readers to existing literature for more specific 4D concepts and algorithms (e.g. \cite{mcmullen2008visual,chu2009gl4d, bosch2020n}).

\subsection{Projection}
Projection is the main method we use to reduce dimensionality of 4D objects and render them on screen. Our approach involves storing 4D geometry data in dedicated memory buffers as \textit{tetrahedral} meshes, and applying a custom projection to obtain associated \textit{triangle} meshes. At this point, objects can be fed through the standard 3D graphics pipeline and rendered on screen. This approach implies the existence of two independent viewers (or cameras): 1) our custom 4D projector, which transforms the input 4D geometry into 3D meshes, and 2) the standard 3D view frustum projection, which enables rendering 3D scenes on screen and is already available in Unity 3D.
We enable the 4D to 3D projection step to be performed in two different ways: \textit{cross-section} and \textit{frustum projection}.

\paragraph*{Cross-section.} This method slices 4D objects based on their intersection with an arbitrary hyperplane. 
Let's consider as a 3D analogy the polygonal cross-section generated by the intersection of an arbitrary plane with a cube. Each edge of the polygon generated by such intersection lies in one of the six faces bounding the cube (which is why the polygon can have a maximum of six sides). Similarly, the intersection of a hyperplane with a tesseract generates a 3D polyhedron, whose polygonal faces each lie in one of the tesseract's eight bounding cubes (Fig.~\ref{fig:projection}). Our cross-section implementation follows the same principle by finding the intersection of a hyperplane with the 4D geometry, defined using the tetrahedron (3-simplex) as a building block. Each intersecting tetrahedron results in a triangle or quadrilateral, and represents one face of the projected 3D mesh. If the intersection is empty, the projection will not produce any 3D output. In \engine ~we define the hyperplane used in the cross-section through a 4D pose (translation + orientation), so that it is straightforward to associate it with the concept of a ``4D camera''.

\paragraph*{Frustum projection (or perspective).} This type of projection forces objects to appear smaller as they get further away from the viewer along the \textit{w} axis. If we visualize the frustum projection of a tesseract, we notice a second, smaller cube inside the larger cube that we would not normally observe in cross-section. This second cube is one of the eight bounding the tesseract, and it appears smaller because it lies further away along the $w$ axis. We note how the 2D rendering of the tesseract in Fig.~\ref{fig:projection} has two distinct vanishing points, which are associated with our 4D camera and the standard 3D camera, respectively. While outside the scope of this paper, it is worth reminding that frustum projections of certain 4D geometry may self-intersect \cite{zachariavs2000projection}. \engine ~does not directly address such cases, since their implications on the user experience vary based on the type of application.


\subsection{Transformations}

Below we describe how standard transformations such as translation, rotation and scaling generalize to four dimensions, and which perceived effects these transformations create on the part of the viewer.

\paragraph*{Translation.}
In cross-section, a 4D object moving along $w$ can be differently sliced based on its intersection with the current hyperplane, until all of its vertices exit the hyperplane, making the object itself invisible to the eye. In Fig.~\ref{fig:bullet}, for instance, an imaginary weapon shoots a bullet in the shape of a hyperdodecahedron (also known as a ``120-cell''), enforcing upon such bullet a sinusoidal movement along $w$. The effect perceived by the viewer is that of a morphing geometry that periodically appears on and disappears.
In frustum projection, an object moving further away along $w$ is perceived as smaller, while an object moving closer becomes larger. If the 4D viewing frustum and the standard 3D frustum do not share the same $x$, $y$, and $z$ coordinates, the object can be perceived as increasingly warped the closer it gets to the 4D camera (examples of this effect can be seen later in Fig.~\ref{fig:warping} and Fig.~\ref{fig:dragon}).

\begin{figure}[t]
\centering
 \includegraphics[width=1\linewidth]{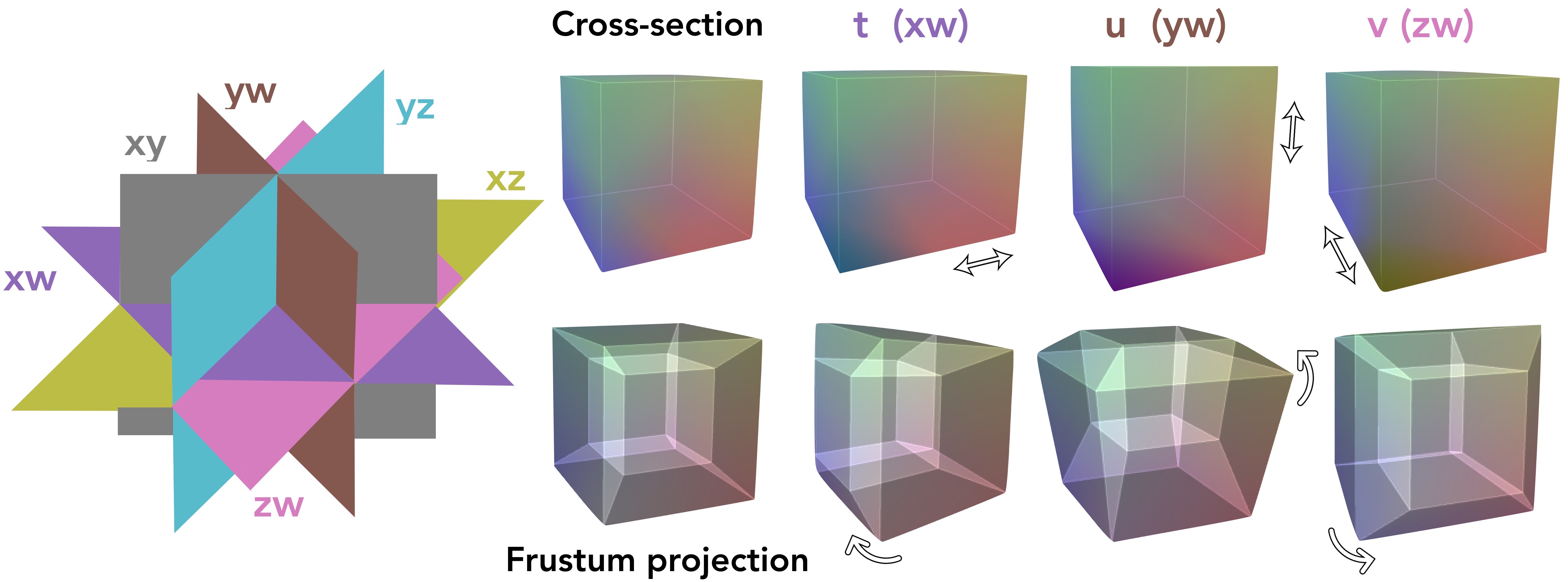}
   \vspace{-1.5em}
 \caption{\textbf{4D rotations}.
 Rotations in four dimensions do not just have an axis, but a plane around which everything else rotates. The four orthogonal axes define four mutually orthogonal 3D hyperplanes or \textbf{six} 2D orthogonal planes. For this reason, \engine ~defines rotations in 4D space through a set of six angles (three variables---$t$, $u$, and $v$---more than a conventional 3D rendering engine). Standard 3D rotations exhibit the expected behavior, but rotations around the three new planes are perceived as mesh deformations. The figure above shows how rotating a tesseract seems to stretch the mesh in cross-section, and to turn the object inside-out when using frustum projection.}
 \label{fig:rotation}
 \vspace{-1em}
\end{figure}

\paragraph*{Rotation.}
Rotations in four dimensions do not generalize to an additional rotation around the $w$ axis. The number of \textit{principal rotations}, in fact, is given by the number of distinct \textit{pairs} of orthogonal axes. As shown in Fig.~\ref{fig:rotation}, the four axes generate \textbf{six} orthogonal planes, which we denote as x, y, z, t, u, and v for brevity (with t, u, and v indicating the xw, yw, and zw planes, respectively). Rotation around the first three can be referred to as the 3D rotation of a 4D object, and does not reveal any information about the higher geometry of the object. Meanwhile, rotating a tesseract around the t, u, v planes produces an interesting visual effect: in cross-section, the 3D projection of the tesseract seems to stretch in one direction; in frustum projection, the cubes composing the tesseract seem to ``turn inside-out'' (Fig.~\ref{fig:rotation}). While it is convenient to express rotations through the six variables described above, \engine ~internally represents rotations as 4x4 matrices (and general transformations as 5x5 matrices) to avoid gimbal lock \cite{aguilera2004general,sakai2011fourb}.

\begin{figure}[t]
\centering
 \includegraphics[width=0.9\linewidth]{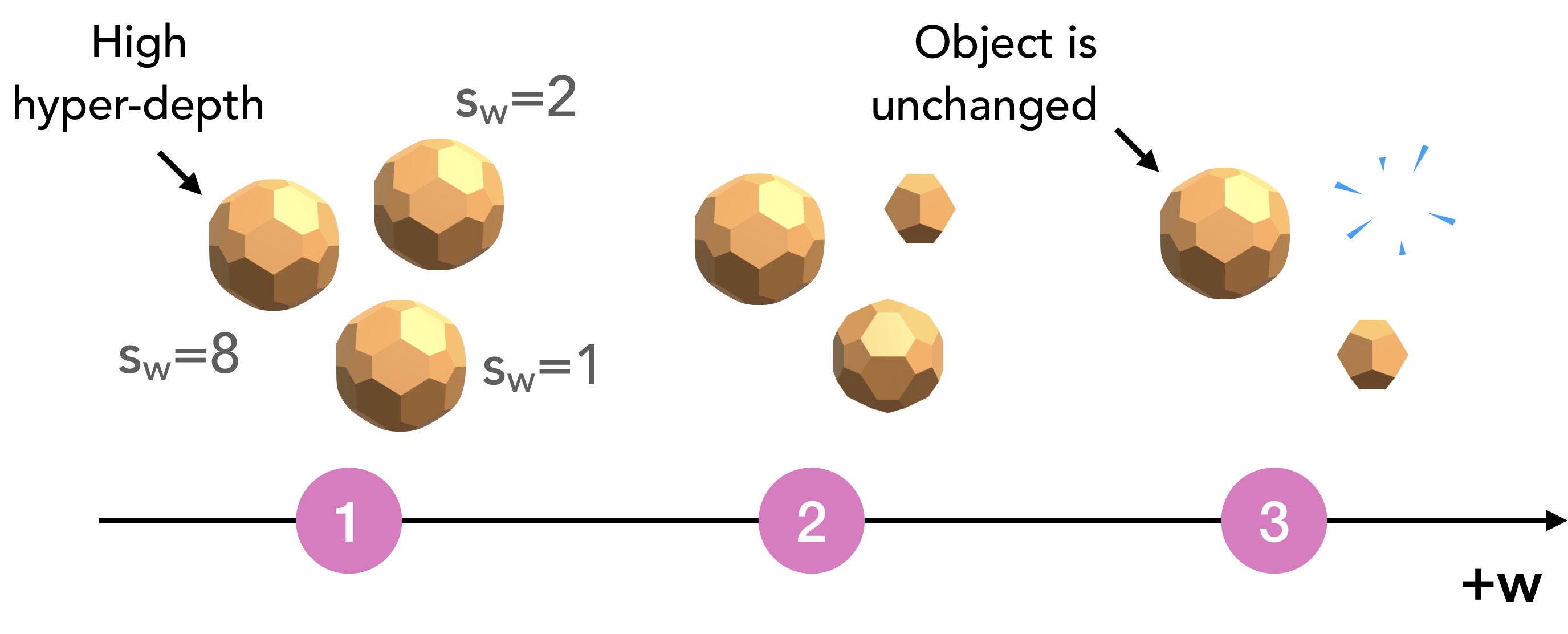}
   \vspace{-0.5em}
 \caption{\textbf{Offsetting hyper-depth}. Scaling an object along the fourth dimension changes the $w$ coordinate of its vertices. This can influence the size and the geometry of its projection. The image above shows cross-sections of three identical---but differently scaled---4D objects as the viewer moves in the $w$ direction. We note how projections of objects with higher hyper-depth are more \textit{persistent}.}
 \label{fig:hyper-depth}
 \vspace{-1em}
\end{figure}

\paragraph*{Scale.} When using the cross-section approach, we can imagine the extent of an object in the fourth dimension (``hyper-depth'') as its resistance to translations along $w$. As shown in Fig.~\ref{fig:hyper-depth}, the projection of larger scale objects \textit{persists} (i.e. is visible) for a longer period of time as the viewer moves along $w$. In frustum projection, scaling a 4D object means its vertices move closer or further away from the viewer, creating a 4D analogue to standard depth. For instance, increasing the scale of a tesseract would make its ``inner'' cube (the one situated at $w=1$) appear smaller and smaller to the 4D viewer.

\subsection{Physics}
\begin{figure}
\centering
 \includegraphics[width=0.9\linewidth]{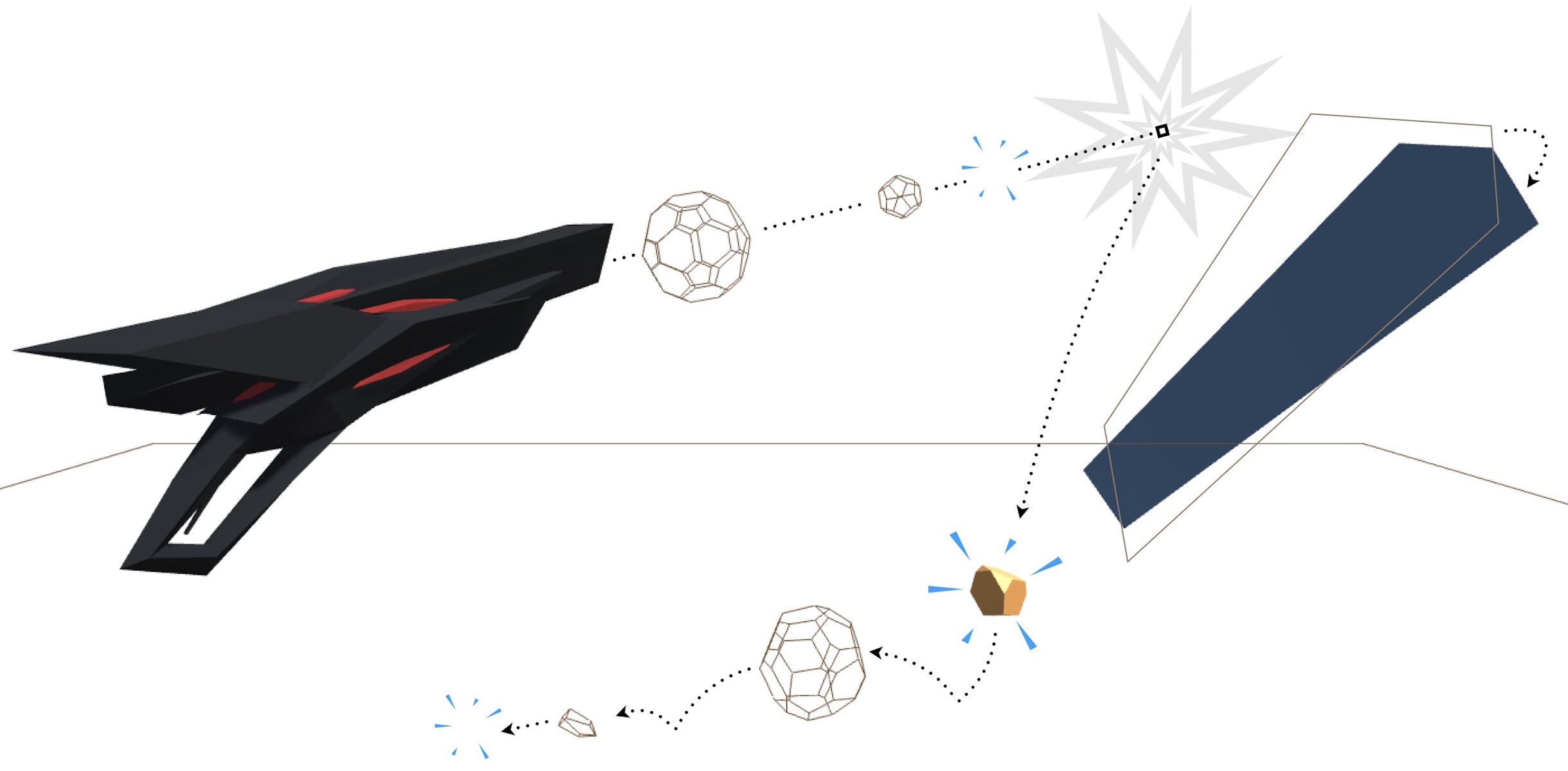}
   \vspace{-0.5em}
 \caption{\textbf{4D Physics.} In \engine, physics on 4D objects are treated as straightforward generalizations of 3D physics. In the example above, the same gun from Fig.~\ref{fig:bullet} shoots a bullet which exits the current cross-section and hits a hidden hyperplane. As result of the collision, the hyperplane slowly rotates around the $t$, $u$, and $v$ planes and becomes partially visible, while the bullet falls to the ground due to gravity, rolling away until its $w$ coordinate again falls outside of the cross-section.}
 \label{fig:physics}
 \vspace{-1.5em}
\end{figure}

\engine ~implements a straight generalization of the laws observed in the 3D universe, assuming principles such as Newton's universal gravitation still hold for higher dimensional objects. Similarly to 4D Toys \cite{bosch2020n}, we support a simple collision system using hypercubes and hyperspheres as colliders, enabling interaction between 4D objects such as the one shown in  Fig.~\ref{fig:physics}. While this is not the central topic of this paper, we want to note that higher dimensional physics may not necessarily generalize our existing understanding of physics in the 3D world. For instance, Newton's universal gravitation and Coulomb's electromagnetic laws follow from Gauss's law, which (assuming its validity in higher dimensions) would change from an inverse-square to an inverse-cube law due to the extra degree of freedom in the fourth dimension \cite{mcmullen2008physics}. This change would obviously affect how object trajectories, spin, and energy are computed.

%% file: method_2.tex
\section{Conceiving a 4D Universe}

In this section we go beyond the scope of visualizing simple 4D phenomena and ask ourselves which laws, constraints, and design criteria should broadly be kept in consideration when designing 4D applications. At the same time, we explore different content and authoring patterns to facilitate the editing of 4D scenes.
A subset of the design considerations outlined here will be then operationalized in the use case presented in Section 5.

\subsection{Structure and Movement}
Conceiving an n-dimensional application strictly defines the number of independent directions in which its space extends, but does not necessarily
indicate how objects \textit{move} within such a space. Depending on inherent geometrical structure and dimensions, different movement scenarios are in fact possible.

\paragraph*{Euclidean hyperspace vs compact dimensions.} If we consider a car riding on a roller coaster, or an ant crawling around a metal can, we immediately think of trajectories winding through a three-dimensional world. However, a closer look reveals that the roller coaster ride can be described with only one independent variable, and is effectively one-dimensional. Similarly, the ant will perceive its own movement on the surface of the can as two-dimensional, and would be completely unaware of living in a 3D world \cite{mcmullen2008visual}. In relation to this concept of \textit{compactification}, a four dimensional application does not necessarily imply the use of 4D content nor the ability to freely move in all dimensions. In this sense, we might consider adding spatial constraints within the ``flat'' Euclidean model presented in Section 3. We could imagine, for instance, the existence of 3D beings living in the different 3D subspaces of a tesseract, with these subspaces being connected through curved dimensions such as a wormhole or a 3D passage. Despite living in a higher dimensional space, such beings would only experience a 3D universe, and their movement would be constrained to a lower number of independent variables.


\paragraph*{Ability to move in higher dimensions.} Common experience suggests that humans cannot voluntarily move along a fourth dimension. String theory, however, suggests the existence of subatomic particles, called \textit{gravitons}, that can move in all existing space (and time) dimensions. In computer graphics applications, it might be convenient to imagine the existence of certain macroscopic beings who are like gravitons, and therefore able to move across all dimensions. For instance, such a 4D being would be capable of escaping any 3D prison by moving along $w$, suggesting a scientifically-minded take on fictional concepts such as ghosts being able to pass through walls or disappear.

\paragraph*{Spatial extent of higher dimensions.} Astrophysical and cosmological observations, high-energy particle collider data, and experiments on Newton's law of universal gravitation \cite{mcmullen2008physics} suggest that any extra dimension would have to be bounded in space to explain the limits of human perception.
According to superstring theory, such extra dimensions would be sized in order of Plank length (about $10^{-33}$cm)---which would be highly impractical in a concrete 4D application, as it would eliminate 4D perspective and cause abrupt hyper-plane transitions in cross-section. However, it might be worth considering a range of admissible $w$ coordinates for facilitating the authoring process, for defining multiple gravity vectors, or for performance reasons. In cross-section, for instance, user interactions with visible 4D objects can happen only in a small window of $w$ values, proportional to the hyper-depth of 4D content.




In the examples in this paper, we chose to model the universe as a four-dimensional, unbounded, Euclidean (non-compact) space. In such hyper-space, only certain macroscopic beings can move in the fourth dimension, along which we do not consider any gravitational force. While not compliant with physical evidence about the structure and extent of higher dimensions, this implementation is (almost) a straightforward generalization of 3D Euclidean space, and its practicality makes it convenient for studying the effects of 4D projections.

\begin{figure}[t]
\centering
 \includegraphics[width=1\linewidth]{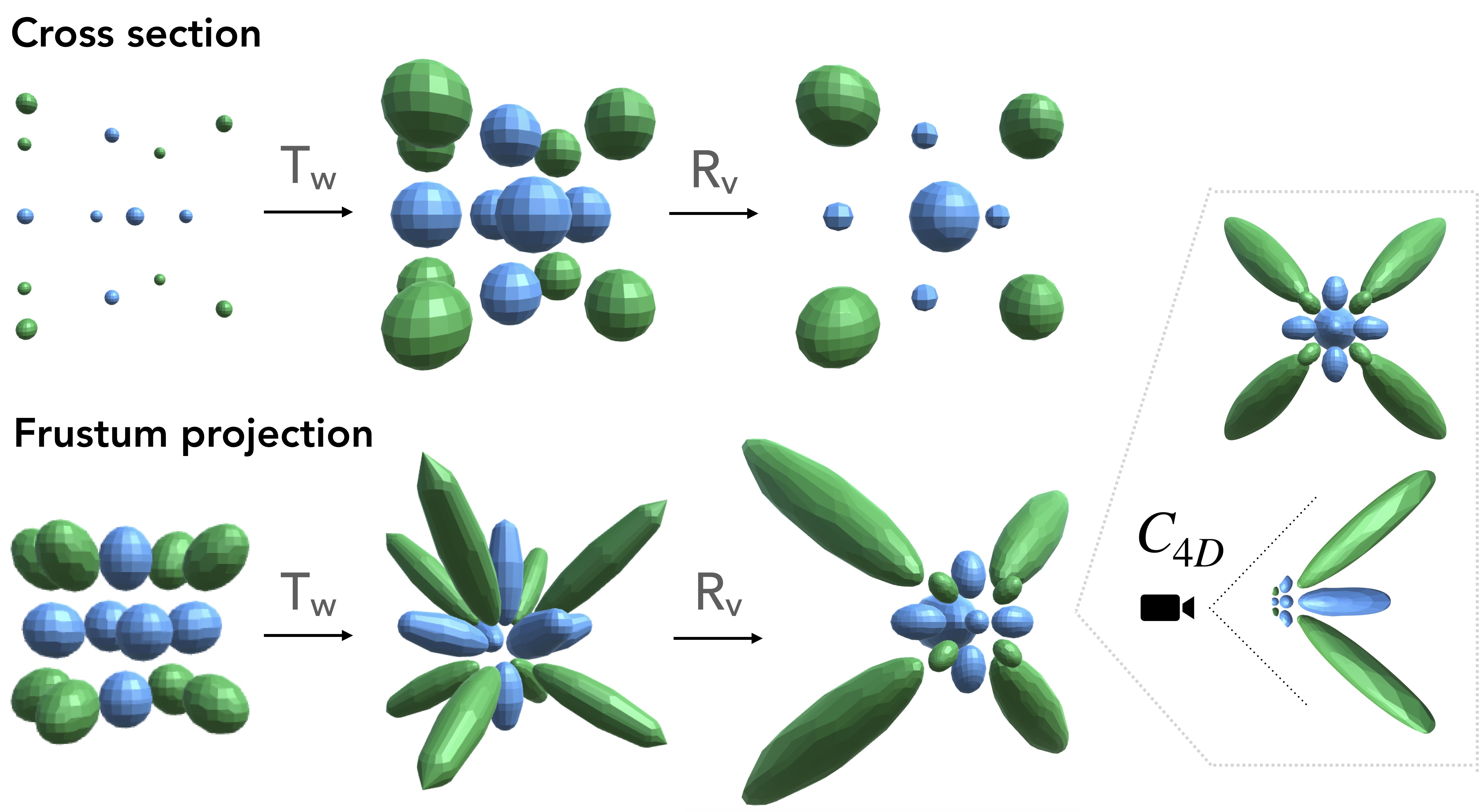}
   \vspace{-0.5em}
 \caption{\textbf{Animating 4D primitives.} This figure shows the perceived shape deformation associated with the translation of ten hyperspheres along $w$, followed by a rotation around $v$. This process demonstrates how simple 4D primitives can be combined and animated to form non-trivial, dynamic 3D projections. We note how, in frustum projection, meshes stretch along the direction of the 4D viewing frustum based on their $w$ distance.}
 \label{fig:warping}
 \vspace{-1em}
\end{figure}

\subsection{Content}

After defining the structure of our hypothetical hyperuniverse, the next step is to try to identify the entities who might populate it, and how they could be constructed.

\paragraph*{Procedural generation through mathematical functions.} Similarly to how a tesseract can be created by extending a cube along the $w$ axis, several 4-polytopes (also known as \textit{polychora}) can be constructed from regular polyhedra \cite{coxeter1973regular} using \textit{lift} or \textit{join} operations \cite{brisson1993representing}. Just as a 3D polyhedron is bounded by polygonal faces, so a polychoron is bounded by polyhedral cells, which need to  meet under certain angle constraints in order to form a closed 3-surface \cite{briggsexploring}. 
These perfect geometries are fundamental for educational purposes and can be used as 4D primitives, sometimes delivering surprising visual effects like the one shown in Fig.~\ref{fig:warping} or like the crystal geometries in Miegakure \cite{miegakure}. However, 4D primitives tend to fall short of conveying the complexity and realism of the world we are used to living in.

\paragraph*{Analogy-based design.} 4D content can be designed by observing how certain concepts generalize from lower to higher dimensions. To start, we could try to imagine the nature of imaginary 4D humanoids by looking at how 2D and 3D beings behave, and then considering the objects and machinery they might build while trying to generalize devices such as wheels, gears, belts and pulleys \cite{mcmullen2008visual}.
For instance, we can imagine a 2D stool as having at least one leg---or more likely two legs---to avoid falling over. In the 3D world, a three-legged stool would balance perfectly, since any three non-linear points lie in a plane, while the presence of more than three legs could make the stool wobble if one leg were slightly longer than the others. Thinking along these lines, we can imagine a 4D stool as a cubinder with at least four legs, and placed next to one of the six cuboid-shaped sides of a 4D table. The same principle of balance would have us consider 4D humanoids to possess at least three legs---possibly four for stability and symmetry. 
Straightforward generalization is sometimes challenging and not always possible. For instance, in 3D we can connect objects together with a rope, which needs to be tied in a knot. This operation is not feasible in 2D, nor is it in 4D, where the knot would easily come undone due to the extra degree of freedom. In relation to the mathematical concept of \textit{chirality}, humans in 3D space cannot superimpose their left hands onto their right hands. This operation is possible, however, through rotations in 4D \cite{dalvit2012see}.
Higher dimensional objects may also not have direct 3D analogies, and be conceivable only by hypothetical 4D beings. For instance, it would be possible for 4D beings to paint 3D pictures without the need for perspective, or, like humans can spot a cavity in a 2D surface, to directly observe the internal structure of 3D objects (theme known as ``clairvoyance'' \cite{leadbeater1903clairvoyance,bragdon1923primer}). 
While many other lower-to-higher dimensional analogies can be found in the literature \cite{abbot1884flatland,rucker2014fourth,rucker2002spaceland}, this design methodology requires systematic thinking and is not easily scalable---making it difficult to apply beyond simple prototypes.

\paragraph*{Data-driven shape synthesis.} Another way of creating 4D content could stem from existing research on 3D \textit{shape synthesis}, which aims to build generative models to support the creation of new, diverse, and realistic shapes along with associated part semantics and structure.
Many approaches to synthesizing geometry involve 3D reconstruction, which uses lower-dimensional features, visual clues and prior information about geometry to generate 3D shapes.
Examples include the estimation of 3D human joint positions and skin surfaces from 2D images \cite{guler2019holopose}, and the sketch-based generation of freeforms from sparse 2D strokes \cite{li2017bendsketch}.
An alternate approach involves the synthesis of objects from a latent lower dimensional vector \cite{hinton1990connectionist}. This generally requires an initial dataset of 3D geometries encoded into a lower-dimensional embedding, which is then used to generate variations on such geometries \cite{li2017grass,wu2018structure,mo2019structurenet,chang2015shapenet}.
While promising, these approaches still deal with simple and mostly convex geometry, represented as voxels (point clouds) or bounding boxes, and are still far from being generalizable. Applying such methods to 4D content generation has  two primary limitations: 1) since we do not know how 4D objects might actually look and behave, it is hard to define semantically meaningful geometrical constraints, and 2) it is not currently feasible to build a curated dataset of 4D objects that would contain enough data samples and variation to allow for a machine learning approach.

\begin{figure}
\centering
 \includegraphics[width=\linewidth]{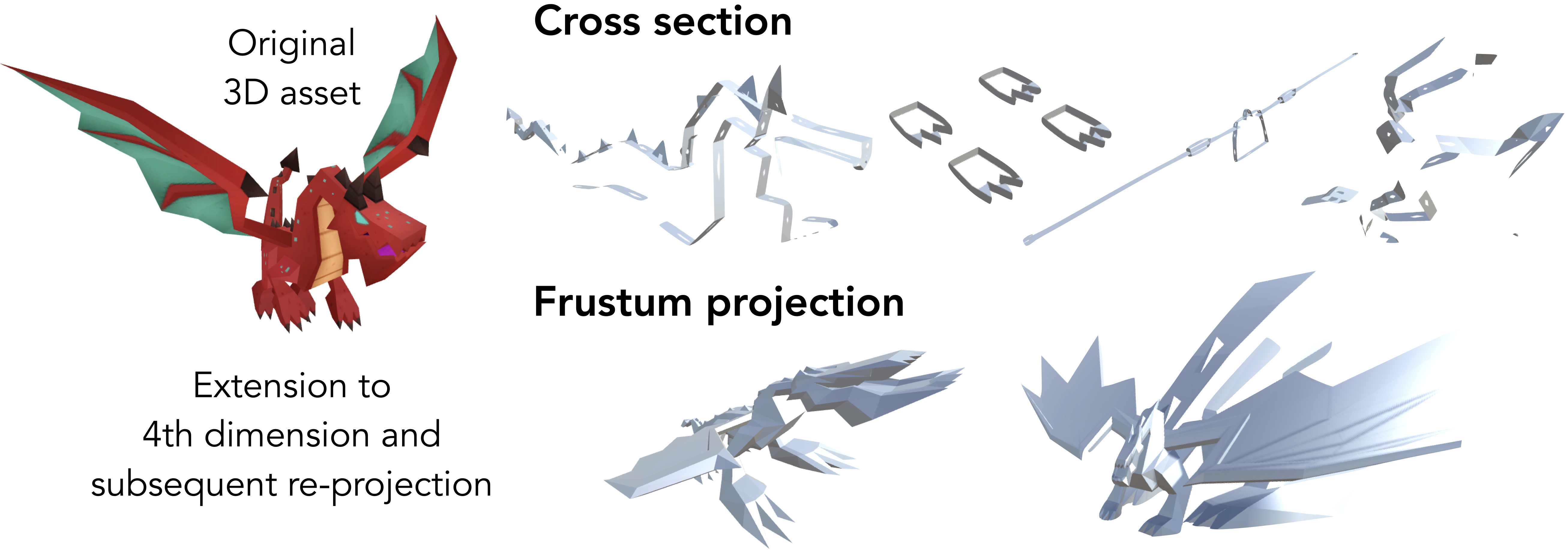}
   \vspace{-0.5em}
 \caption{\textbf{Reprojecting existing 3D assets.} As well as  4D polytopes, standard 3D meshes can also be handled in four dimensions. In the example above, a 3D asset is extended along $w$ and then reprojected into 3D. We notice how 4D rototranslations slice the mesh into unexpected shapes in cross-section, whereas frustum projection shows the mesh stretching in the direction of the 4D camera.}
 \label{fig:dragon}
 \vspace{-1em}
\end{figure}

\paragraph*{Extension and 4D transformation of existing 3D assets.} A final, more simplistic approach, consists of extending existing 3D objects into the fourth dimension, similarly to how we extend a cube into a tesseract. Specifically, we use lifting and tetrahedralization \cite{bern2003adaptive} in the examples shown in this paper.
While likely sacrificing the higher dimensional significance of such objects, this approach has the advantage of guaranteeing 4D object reprojections appear similar to their 3D counterparts when at rest. By applying simple 4D transformations, we can tune the 3D appearance (as shown in Fig.~\ref{fig:dragon}), all while retaining most of the semantics associated with a three-dimensional experience. This approach can be interpreted as a method to \textit{use 4D to generate new 3D content}, more than as a way to create meaningful 4D content.

\begin{figure}[t!]
\centering
 \includegraphics[width=0.8\linewidth]{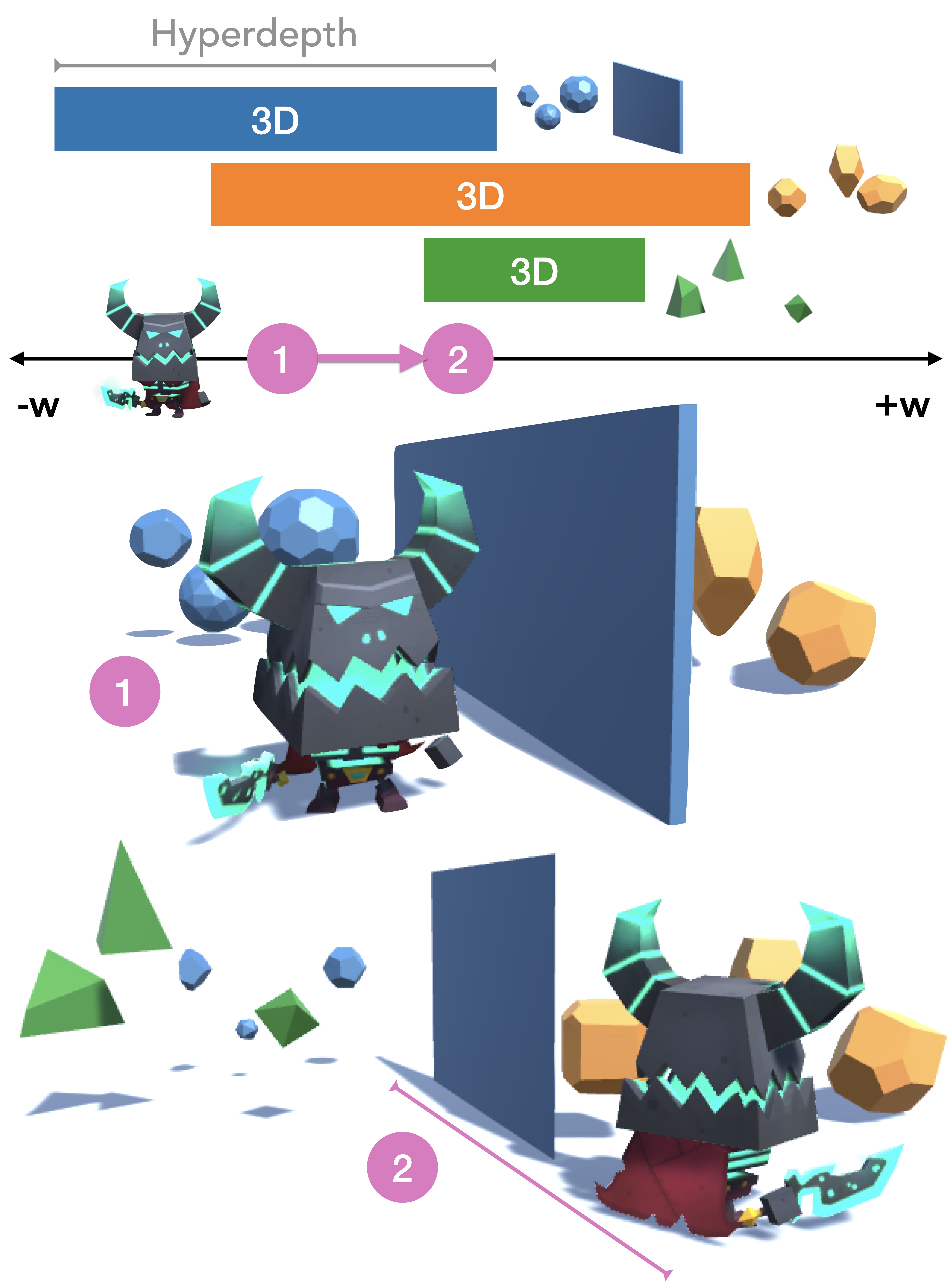}
   \vspace{-0.5em}
 \caption{\textbf{Combining 3D worlds by offsetting their hyper-depth.} One possible way to author a 4D environment consists of designing a set of 3D environments, extending them along the fourth dimension, and then placing them at different $w$ coordinates in 4D space. A viewer moving along the fourth dimension would experience a smooth transition from one 3D world to the next. Objects from different 3D environments can be simultaneously visible, with possibly different \textit{persistence}, based on the overlap of their hyper-depths. In the cross-section example above, a character cannot reach its target due to the presence of a wall belonging to the \textit{blue} world (1). By applying a translation along $w$, the wall shrinks while exiting the cross-section, allowing the character to pass (2). At the same time, objects from the \textit{green} world become visible.}
 \label{fig:overlay}
 \vspace{-2em}
\end{figure}

\begin{figure}[t!]
\centering
 \includegraphics[width=0.9\linewidth]{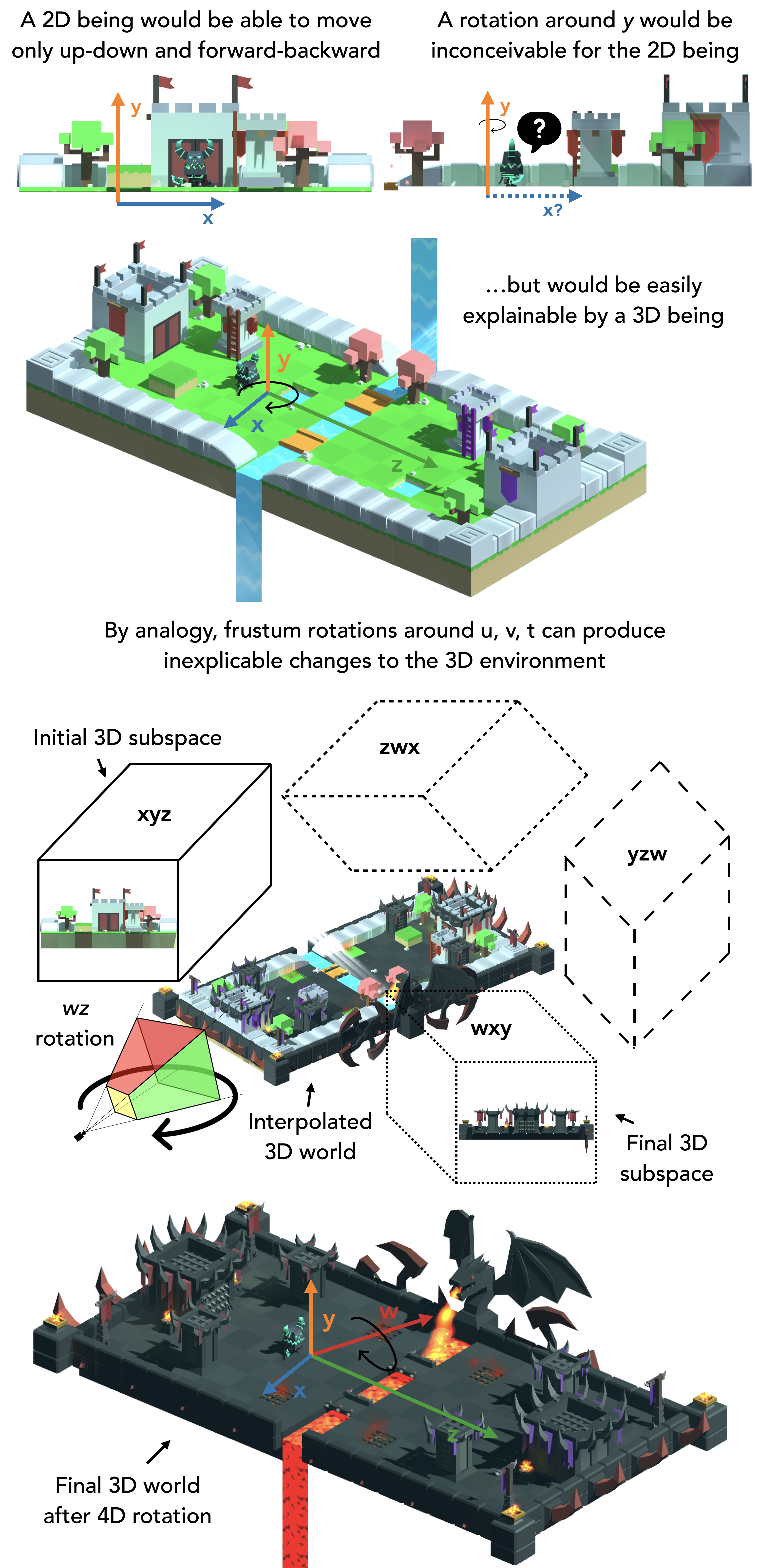}
   \vspace{-0.5em}
 \caption{\textbf{Combining 3D worlds through orthogonal subspaces.} A possible technique to adopt when authoring 4D environments consists of using 4D rotations to transition between different 3D worlds. To use an analogy, a two-dimensional being, unaware of the fact that it lives within a 3D world, would only be able to see two directions of motion, and could rotate only around the $z$ axis, perpendicular to the $xy$ plane it lies on. If this character could perform the action---to him inconceivable---of rotating around $y$, the world surrounding him would inexplicably change. This change, however, would make perfect sense from the perspective of someone able to perceive the higher dimension. Similarly, 4D rotations would enable the viewer to transition from one 3D world to another, with the effect of ``warping'' the surrounding environment.}
 \label{fig:subspaces}
 \vspace{-2em}
\end{figure}

\subsection{Authoring}

After having defined the fundamental structure of our hypothetical hyperuniverse and the possible entities inhabiting it, the next challenge lies in identifying methods and design patterns to concretely author meaningful 4D experiences. While 3D editing tools such as Unity 3D, Unreal Engine, and Autodesk Maya have now become ubiquitous, there are currently no general-purpose tools dedicated to the creation of higher dimensional environments. Central design challenges include the difficulty of conceiving and manipulating such geometries, differing camera controls, the scarcity of available 4D content, and the integration of 4D content with existing 3D assets. In the following, we assume the usage of a standard 3D editor (e.g. Unity3D) with \engine ~integration as authoring tool---implying that the user is able to translate, rotate, and scale objects in three dimensions as usual (e.g. through gizmos and mouse interaction, or manual numerical input). In this context, the $w$ coordinate, 4D rotations, and scaling of a 4D object may be arbitrarily modified by the user. Doing so will generate the projection of a 4D object, which is then treated as a normal 3D asset within the authoring tool. From the user's perspective, this makes manipulating 4D objects similar to handling 3D meshes, with the additional possibility of regulating their geometry and visibility through a set of \textit{4D parameters}.




\paragraph*{Layout.} One design choice involves deciding how much \textit{native} 4D content to include and how to position it in hyperspace.
A possibility would be to consider authoring an environment where \textit{all} content comes from intrinsically 4D geometries.
Unfortunately, most available 4D content derives from abstract mathematical entities whose natural interactions in higher dimensions would be particularly hard to conceive of within a 3D editor. In this case, proper placement and behavior of such objects could be defined through scripting and custom mathematical functions, and then visually validated through the authoring tool---but the result would be a geometrical world probably too different from anything resembling the common human experience. \\
An alternative approach would be to let the user author 3D worlds as usual, and then combine them within the four-dimensional space according to some \textit{placement} criteria. One simple example would involve defining a set of independent 3D worlds, and lending each of them a different coordinate along the $w$ axis. If we assume such worlds have no hyper-depth, a viewer would be able to experience those worlds, one at a time, when moving at their respective $w$ coordinate. However, if we extend the 3D worlds into 4D so as to arbitrarily define their hyper-depth, we can create a \textit{transition} effect between such 3D environments, possibly overlapping their content and creating environmental effects such as doors opening or objects materializing (Fig.~\ref{fig:overlay}).
In a similar fashion, 4D rotations can be used to transition between 3D worlds placed in different subspaces of our hypothetical hyperuniverse (Fig.~\ref{fig:subspaces}). 
Such transitions would require, however, extending any existing 3D asset into four dimensions, and fully recomputing the entire scene geometry---operations that can significantly reduce the application performance, as later discussed in Section 6. Possible workarounds include renouncing some 4D transition effects (perhaps ``simulating'' them through a standard 3D engine), and simplifying the scene geometry. Miegakure \cite{miegakure}, for instance, achieves smooth cross-section performance by limiting the number of visible objects and emphasizing the of use of textures on simple 4D primitives.\\
A third and final approach to authoring a 4D scene consists of designing a single, standard 3D world, and integrating a relatively small, but more carefully chosen, number of 4D entities. In this case, intrinsically 3D assets would be unaffected by 4D physics and projection parameter changes. One advantage of this approach is that it does not require us to extend 3D assets into 4D in order to place them in higher-dimensional space, and therefore drastically reduces the computation required to project them back into 3D. At the same time, it allows us to maintain standard 3D authoring practices, focusing on the definition of few selected 4D interactions, which are inherently more time-consuming to design.


\begin{figure}
\centering
 \includegraphics[width=1\linewidth]{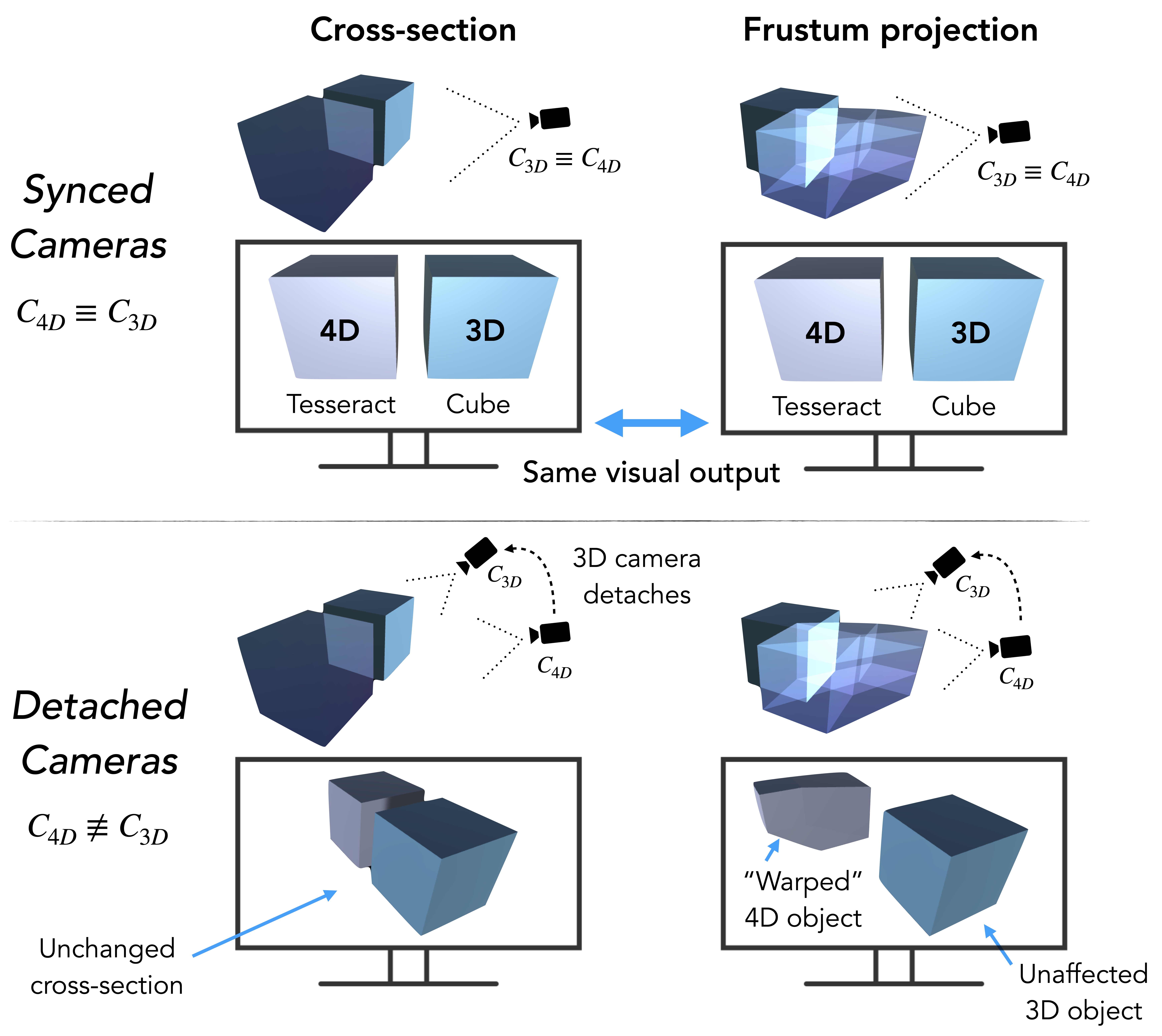}
 \caption{\textbf{Camera setup for mixed 3D/4D environments.} In both cross-section and frustum projection, the 4D and standard 3D cameras can be \textit{synced} as to share the same 3D rototranslation components, and treated as a single viewer. While both projection methods might lead to the same visual output on the screen, the internal 3D representation of the scene generated by frustum projection involves morphed geometries, so as to preserve the 4D perspective effect in the final screen rendering. This implies that, if the 3D camera is \textit{detached} from the 4D viewer, 4D geometries will be perceived as ``warped'' and will behave differently from their 3D counterparts.}
 \label{fig:camera-setup}
 \vspace{-2em}
\end{figure}

\paragraph*{Camera Setup.} While in standard graphics pipelines we have the concept of a single rendering camera, which transforms a 3D world into the visuals we experience on our 2D screen, in our \engine ~engine the designer has an additional viewer (or camera) at their disposal---opening up a set of new authoring possibilities.
If we directly generalize the existing 3D approach, we can imagine the existence of a single viewer, able to independently move and rotate in 4D space. If an application involves only 4D assets, movement within the environment can be performed by simply allowing the user to control the 4D camera, and leaving the 3D camera fixed. Problems arise only when the environment is comprised of both 3D and 4D assets.
If we desire a straightforward generalization of a standard 3D viewer, we can define the 3D and 4D cameras as to share the 3D components of their transformations, with the 4D camera able to independently move along $w$ and rotate around $u$, $t$, and $v$. We call this approach ``synced cameras'', as from the user perspective it is equivalent to control a single, unified 4D viewer. We note how in this case the camera transformation is applied twice to 4D objects (first when they are projected into 3D, and then again when projected to screen space), requiring a simple offset correction on such geometries.
A different methodology involves instead considering the two cameras as independent. The typical use case here would be in applications involving limited 4D assets, where the user can freely move the 3D camera in the environment and the 4D camera is separately used to control the appearance of 4D geometries. 
This ``detached cameras'' approach can possibly hew to the same visual results of the ``synced camera'' approach when using the cross-section method, but produces completely different results when using frustum projection. 
In fact, the two detached cameras guarantee the existence of two separate vanishing points in the final screen rendering, enabling the perception of both depth and \textit{hyper-depth}. Specifically, frustum projection morphs meshes with respect to the 4D viewer when passing from higher-dimensional geometry into 3D space, and the independent movement of the 3D camera can be leveraged to observe such \textit{warped} geometries (Fig.~\ref{fig:camera-setup})---a visual effect that could be considered physically unrealistic or artistically interesting depending on the use case. This approach is exemplified in the next section, where a new method for smoothly transitioning between cross-section and frustum projection is also presented.





%% file: usecase.tex
\section{Use Case: \game}

\begin{figure}
\centering
 \includegraphics[width=\linewidth]{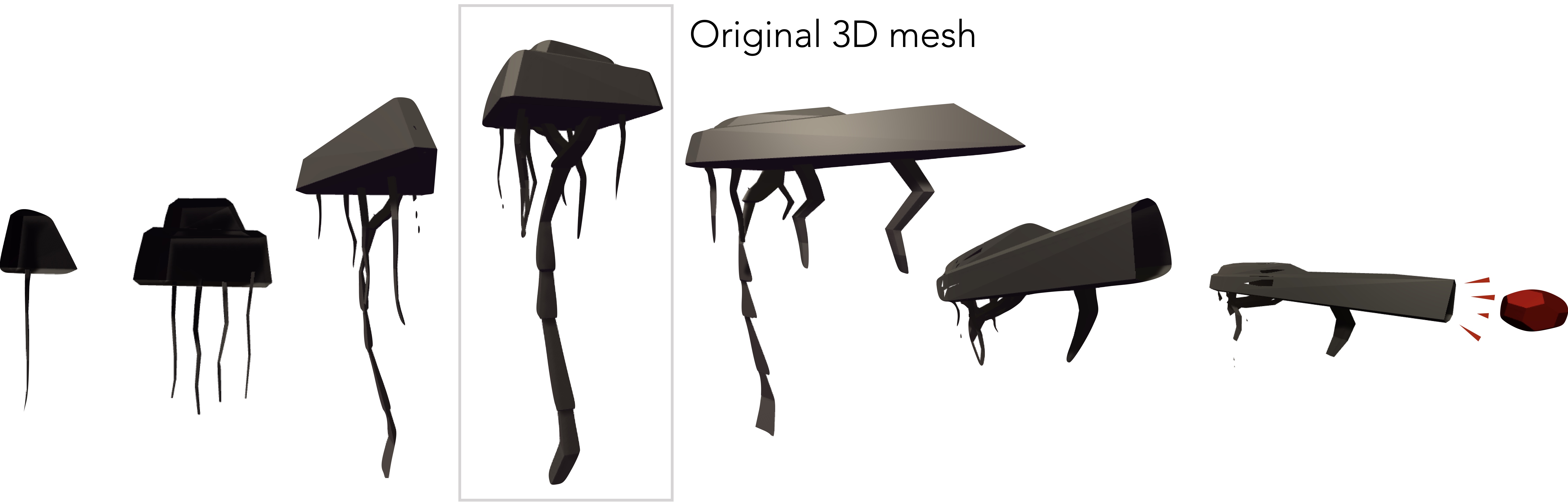}
  \vspace{-1.5em}
 \caption{\textbf{Animating enemies}. After extending existing 3D assets into the 4th dimension, we can animate their 3D reprojections by applying simple 4D transformations. In the cross-section example above, an object morphs from a static tree into a jellyfish, and finally into a gun just by having 4D rotations applied to it.}
 \label{fig:enemy}
 \vspace{-2em}
\end{figure}

In the previous section we outlined a number of  ways to conceive of a hypothetical 4D universe, discussing the potential but also the practicality of each. Here we settle on a \textit{subset} of the methodologies presented in Section 4, and relatedly present the implementation of a prototype single-player videogame called \game.
We designed this game with the specific goals of 1) creating a simple, non-didactic, 4D-enabled application which is not too far from common experiences of 3D (as opposed to existing abstract geometry visualization tools), 2) identifying 4D interactions that are also meaningful from a 3D perspective, and 3) exploring how the cross-section and frustum projection methods can be integrated and organically controlled by the user.
With these criteria in mind, we chose to define a relatively simple hyperuniverse \textit{structure}, characterized by flat, Euclidean geometry (i.e. no compact dimensions). The user is able to freely move along the unbounded $x$ and $z$ dimensions, but is subject to gravity along the $y$ axis and can move along $w$ only within a certain range and under certain circumstances.
So as to preserve familiarity with ordinary 3D videogames and to compensate for the lack of intrinsically 4D assets, the game’s \textit{content} consists of a mix of 4D and 3D objects---and 4D objects exist either as a composition of 4D primitives (Fig.~\ref{fig:warping}) or as the 4D extension of an existing 3D asset (Fig.~\ref{fig:enemy}).
In terms of \textit{authoring}, we started modeling a 3D world using the Unity3D editor, and then made careful additions of hand-picked 4D content using our \engine ~integration, offsetting assets along the $w$ axis based on the principle previously shown in Fig.~\ref{fig:overlay}.
Overall, these design choices had the effect of giving the user the impression of playing a regular videogame at first, and then gradually revealing hidden higher-dimensional content and mechanics.

\begin{figure}
\centering
 \includegraphics[width=\linewidth]{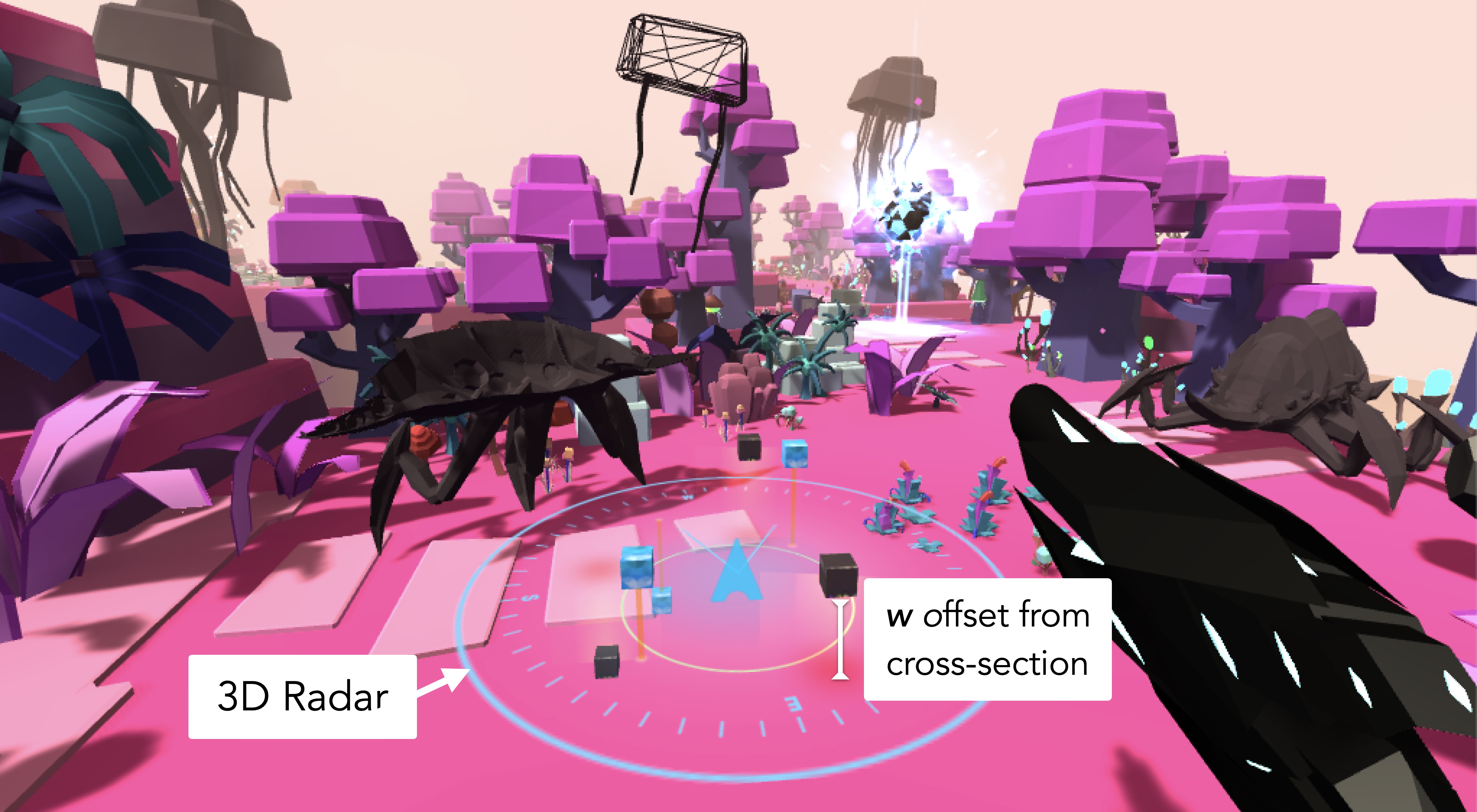}
  \vspace{-1.5em}
 \caption{\textbf{Across Dimensions}. \game ~is a prototype videogame we developed as a use case for our \engine ~engine, where the player can navigate a classic 3D world populated by 4-dimensional beings. The player has at his disposal an arsenal of tools (weapons) that enable him to interact with and manipulate 4D objects, which hide in the environment, and may suddenly appear to confront the player. In the figure above, the player uses 3D radar to estimate the $w$ coordinate of 4D objects and spot enemies despite them being outside of the current cross-section.}
 \label{fig:game}
 \vspace{-1em}
\end{figure}

\paragraph*{Gameplay.}
\game ~is a first-person shooting game with puzzle and exploration elements, set in a fictional sci-fi world. While exploring this apparently 3D world, the player discovers and collects special crystals that fill up a dedicated energy bar, which can be drawn on in order to perform 4D actions such a moving along the $w$ dimension, manipulating higher-dimensional objects, and modifying projection parameters and perspective. Most of these actions are performed through a set of sci-fi weapons with unique abilities. These are also used to fight 4D enemies that the user encounters along the way.
In the current implementation, \game ~is a laptop-based videogame available on Windows and MacOS, and uses standard mouse and keyboard controls as input.

\paragraph*{Navigation and camera control.}

Basic navigation in 3D space is performed through the standard combination of directional keys (WASD or arrow keys) and mouse movement for camera rotation around the $x$, $y$, and $z$ axes. When enough crystal energy is accumulated, the player presses two separate keys to smoothly move in the $w$ direction, which has the effect of allowing him to encounter or leave behind certain 4D objects positioned at different $w$ coordinates. Meanwhile 3D assets in the scene always remain visible.
To help the player navigate a world where many objects are hidden in higher dimensions, we introduce the concept of holographic 3D radar (Fig.~\ref{fig:game}). Similarly to a normal radar display, the user’s location is positioned at the center, with the radar plane defined in the $x$ and $z$ directions, and positioned at the same $w$ coordinate as the player. 4D objects appear on the radar as 3D pins with an altitude proportional to their $w$ offset with respect to the player, suggesting that objects showing on the radar plane are currently visible, while objects above or below the radar plane require the player to move along the $w$ direction in order to reach them.

\begin{figure}[t]
\centering
 \includegraphics[width=1\linewidth]{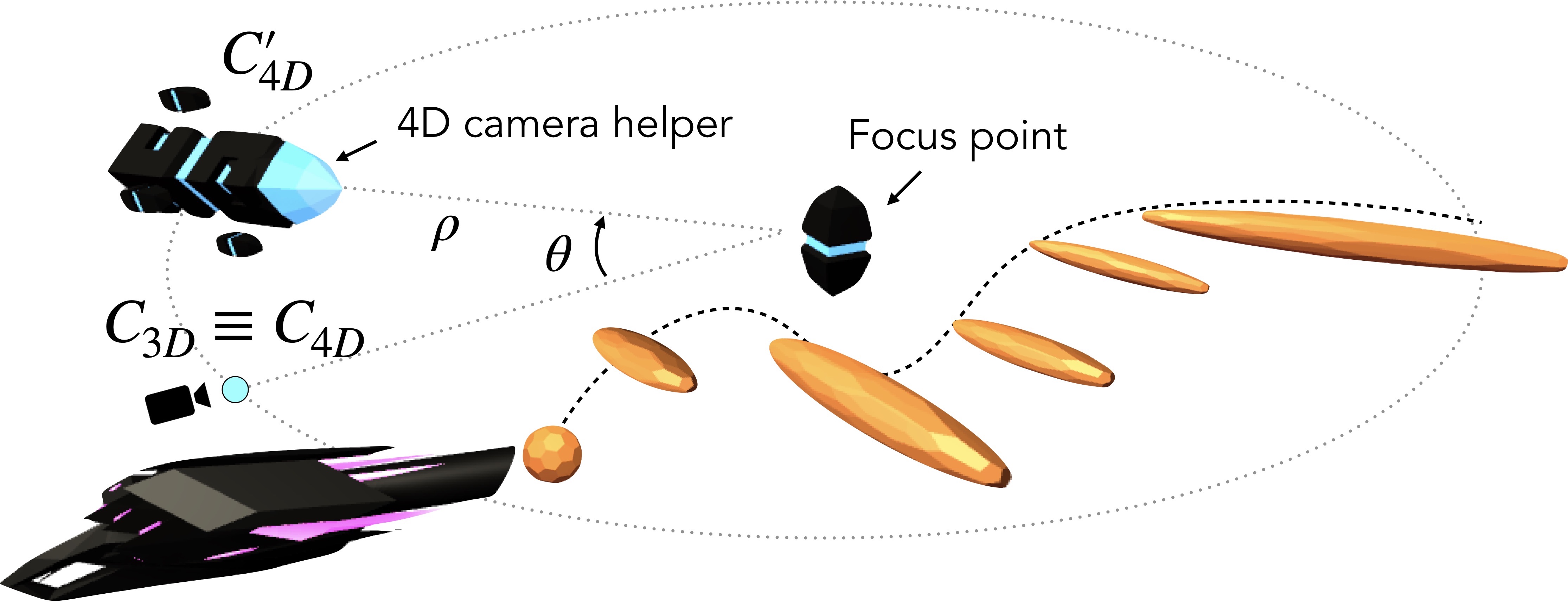}
 \vspace{-1.5em}
 \caption{\textbf{Camera control}. \game ~supports smoothly transitioning from cross-section to frustum projection. When this happens, the player can independently control the 4D camera along an orbital trajectory with the help of in-game visual aids.}
 \label{fig:orbit}
 \vspace{-1em}
\end{figure}

A synced cross-section camera setup is applied by default, so that 4D objects moving along the fourth dimension appear to gradually enter and exit the current cross-section hyperplane. In this modality, the 4D camera simply follows the 3D camera, and is only in charge of changing the cross-section hyperplane. To make the game more challenging and explore novel camera setups, we decided to allow the player to unlock advanced control of the 4D camera.
Using crystal energy, the player can press a key to switch from the cross-section camera to the frustum projection camera.
In the instant in which this transition happens, the 4D camera is in sync with the  3D camera and---if certain criteria are met---no noticeable change is perceived. Specifically, we observed this effect takes place if the surrounding geometry does not have a 4D rotational component, and if there are no nearby 4D objects outside of the current cross-section (as they would otherwise be ``revealed'' by the frustum projection).
When the user moves in 3D space, the 3D camera then detaches from the 4D camera, generating two distinct vanishing points for the $z$ and $w$ dimensions and consequently warping the 4D objects in the scene, as we previously explained in Fig.~\ref{fig:camera-setup}. The warped geometry generated by the frustum projection is a function of the initial position of the user with respect to the 4D object, whose 3D projection then remains unchanged when the user moves around it (since the 4D camera position remains fixed).

The next step involves enabling control of the position and orientation of the 4D camera, which would allow the user to dynamically morph the 4D environment around him after an initial switch to frustum projection. Unfortunately, this operation introduces four new positional and six new rotational variables into the game, and is not easy to conceive of, especially when already focused on the existing mechanics of the game. Therefore, we decided to limit 4D camera control to an orbital trajectory starting from the player’s initial position, and having a fixed focus point around which the camera rotates. Since the focus point is automatically set to be few meters in front of the player during the switch to frustum projection, this leaves only one additional variable, the \textit{orbit angle}, which the user can control via two dedicated keys. Since moving the 4D camera and shifting the $w$ vanishing point are hard to imagine, two new objects appear in the scene when frustum projection is activated, dynamically reflecting the the focus point position and 4D camera position, respectively (Fig.~\ref{fig:orbit}). These two in-game indicators are also shown on the 3D radar display, allowing the player to control them even if they are not currently on screen.\\
When the player decides to switch back to cross-section or when the energy bar is depleted, the 4D camera position gradually rejoins the 3D camera through interpolation, while keeping its orientation locked on the focus point. This method allows for smoothly switching back to cross-section projection with synced cameras, easing the transition between these two completely different projection methods.
While 4D perspective can be used to achieve unique mesh deformations and visual effects, we designed \game ~as to limit the use of frustum projection to a set of carefully chosen scenarios, mostly due to its computational cost and parametric complexity---which require extra time to fine-tune the appearance of 4D geometries and test game mechanics. Cross-section, by contrast, is more straightforward and can be uniformly applied to many 4D objects at a time, enabling the designer to easily create ``hide and reveal'' \cite{miegakure} experiences.

\begin{figure}[t]
\centering
 \includegraphics[width=1\linewidth]{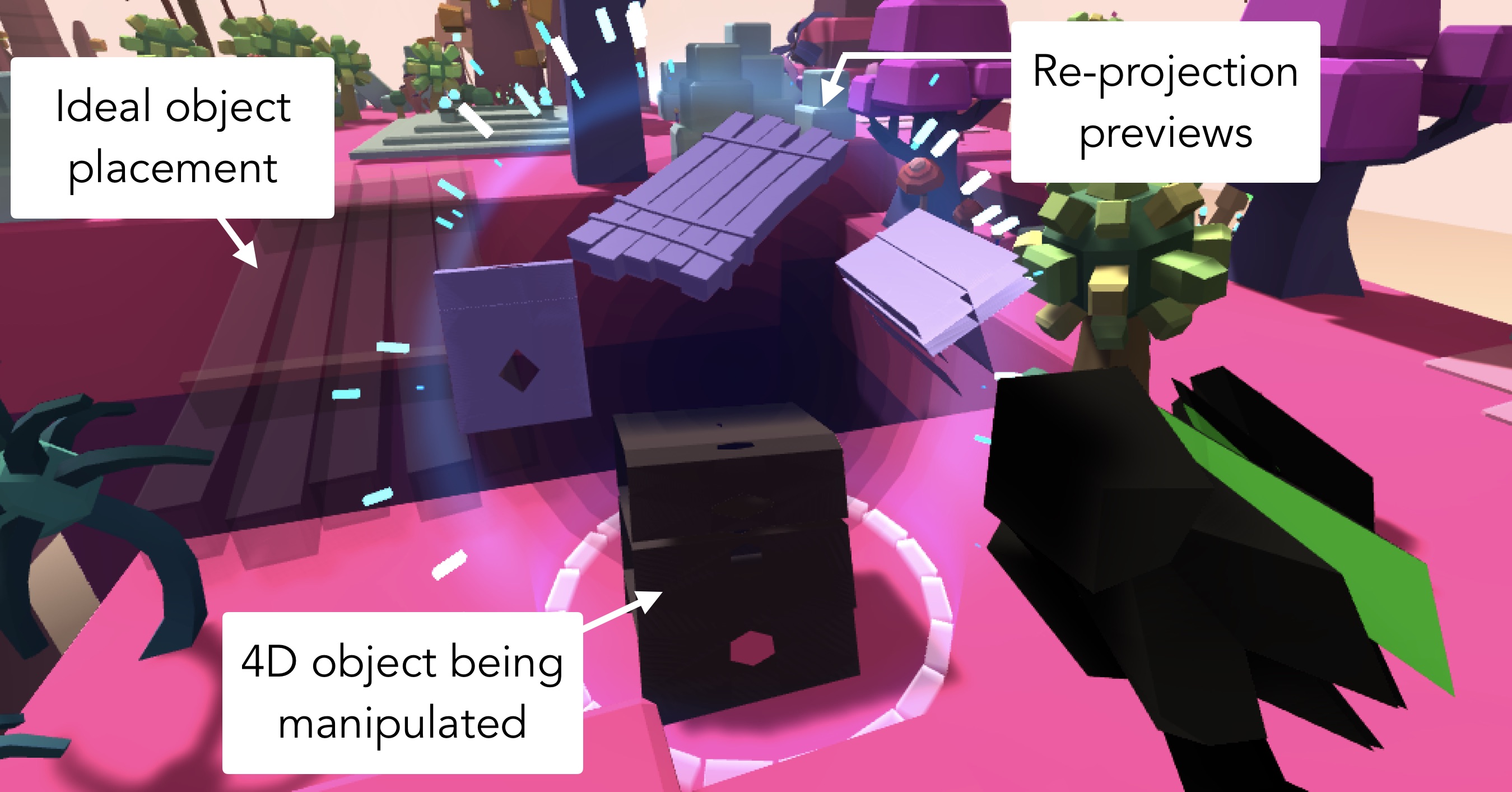}
  \vspace{-1.5em}
 \caption{\textbf{4D manipulation}. To facilitate the manipulation of higher dimensional objects, we preview the effects of 4D transformations on their geometry. In the figure above, the  player applies a 4D rotation to an object in order to build a bridge, creating a path to an adjacent platform.}
 \label{fig:manipulation}
 \vspace{-0em}
\end{figure}

\paragraph*{Interacting with 4D elements.}
Movement  itself represents an initial, basic way of interacting with 4D content. Moving along $w$ while in cross-section can allow the player to pass through walls and reach a target situated behind them (as shown in Fig. \ref{fig:overlay}). In other cases, the movement of the 4D camera can be used to turn certain objects into others, such as turning a barrier into a building that the user can enter to proceed in the game.
When close to certain 4D objects and equipped with a relevant tool, the player can perform direct manipulation of geometry through 4D rotations, utilizing dedicated keys that are indicated on-screen. For example, the player might be stuck on a platform and unable to proceed beyond a steep dropoff. Through 4D manipulation, this problem could be solved by molding a bridge out of a 4D geometry situated at the edge of the cliff, and then placing it so as to create a path to the adjacent platform. Since it is inherently hard to estimate the results produced by 4D rotation, we apply the concept of ``ghost geometries'' \cite{li2015and,kageyama2016visualization} and visualize floating \textit{previews} of the shapes that the 4D object could assume (Fig.~\ref{fig:manipulation}).
The player can also perform more complex actions through a set of weapons which he acquires throughout the game.
One example is the weapon already introduced in Fig.~\ref{fig:bullet} and Fig.~\ref{fig:physics}, which shoots hyperdodecahedral bullets that fluctuate along the $w$ axis. These bullets periodically become invisible as they exit the current cross-section, but their 4D movement can be continuously tracked using the 3D radar display. Each bullet has enough power to move simple 4D objects on contact, allowing one to remove obstacles and hit targets that are currently invisible (e.g. enemies farther away in the $w$ direction. When used in combination with frustum projection, the shape and trajectory of bullets can be dynamically altered by moving the 4D camera, thus making it possible to hit targets around corners or to expand the bullet's collider in a specific direction.
A second weapon has the effect of generating a radial blast that brings all nearby 4D objects to the same $w$ coordinate as the player, revealing hidden objects or enemies. Similarly, another weapon produces an electrical field that modifies the rendering material of nearby 4D objects and shows their wireframe, allowing the user to easily identify them when their presence is disguised among similar-looking 3D assets.

\paragraph*{Fighting enemies.}
Enemies in \game ~are either combinations of 4D primitives (Fig.~\ref{fig:warping}) or are generated by extending existing 3D assets (Fig.~\ref{fig:dragon}), an operation which guarantees content that is visually coherent with its 3D surroundings. Since extending 3D assets does not guarantee the generation of meaningful limbs or skeletal structures in 4D, we decided to ``animate'' enemies simply by applying to their geometry pre-defined 4D rototranslations. This simple method allows for the generation of unconventional visual effects, such as the morphing of a tree into a jellyfish and then into a weapon (Fig.~\ref{fig:enemy}). In \game, enemies hide at different coordinates along the $w$ axis, and can independently move in all four dimensions, freely entering the current cross-section to attack the player or exiting this cross-section to flee. Certain enemies remain partially visible, but hide their weak spots outside of the visible spectrum, so that the player must follow his bullets on the 3D radar after they have exited the cross-section.
Sometimes the player can even loot crystals hidden inside the 4D geometry of an enemy by forcing that enemy to perform particular movements or by controlling the 4D camera.

\section{Discussion}

Here we provide additional considerations that emerged during the \game ~design process, and discuss the challenges and possible benefits of integrating 4D features with existing 3D applications.

\paragraph*{Authoring Hybrid 3D / 4D Worlds.}
The integration of 4D mechanics into \game ~undoubtedly introduced new, creative forms of interaction. Designing the game proved however challenging, particularly with respect to content authoring.
The decision to start by defining a basic 3D environment, and then later add 4D content definitely helped with spatial orientation and facilitated  storytelling. This decision was made due to our desire to not create a completely abstract game, and also to the limited availability of 4D content.
Even when using simple geometries, it is difficult to deploy too many 4D objects in a scene due to an inherent behavioral complexity, which requires that the designer take more time to carefully integrate these objects in the game's storyline. The effects of 4D transformations and camera projections are in fact hard to predict, often leading to a trial-and-error approach to content authoring. In this sense, it may be worth exploring recommender systems for placing and manipulating 4D content, so as to help the designer preview and apply a set of automatically generated 4D behaviors.
Another factor to consider is the trade-off between integrating native 4D elements and approximating some of their behaviors through more conventional techniques such as 3D mesh deformations, which could provide similar visuals with a more deterministic behavior. Along these lines, we recommend a further study to explore which mechanics can be achieved uniquely through 4D interactions.

\begin{figure}
\centering
 \includegraphics[width=\linewidth]{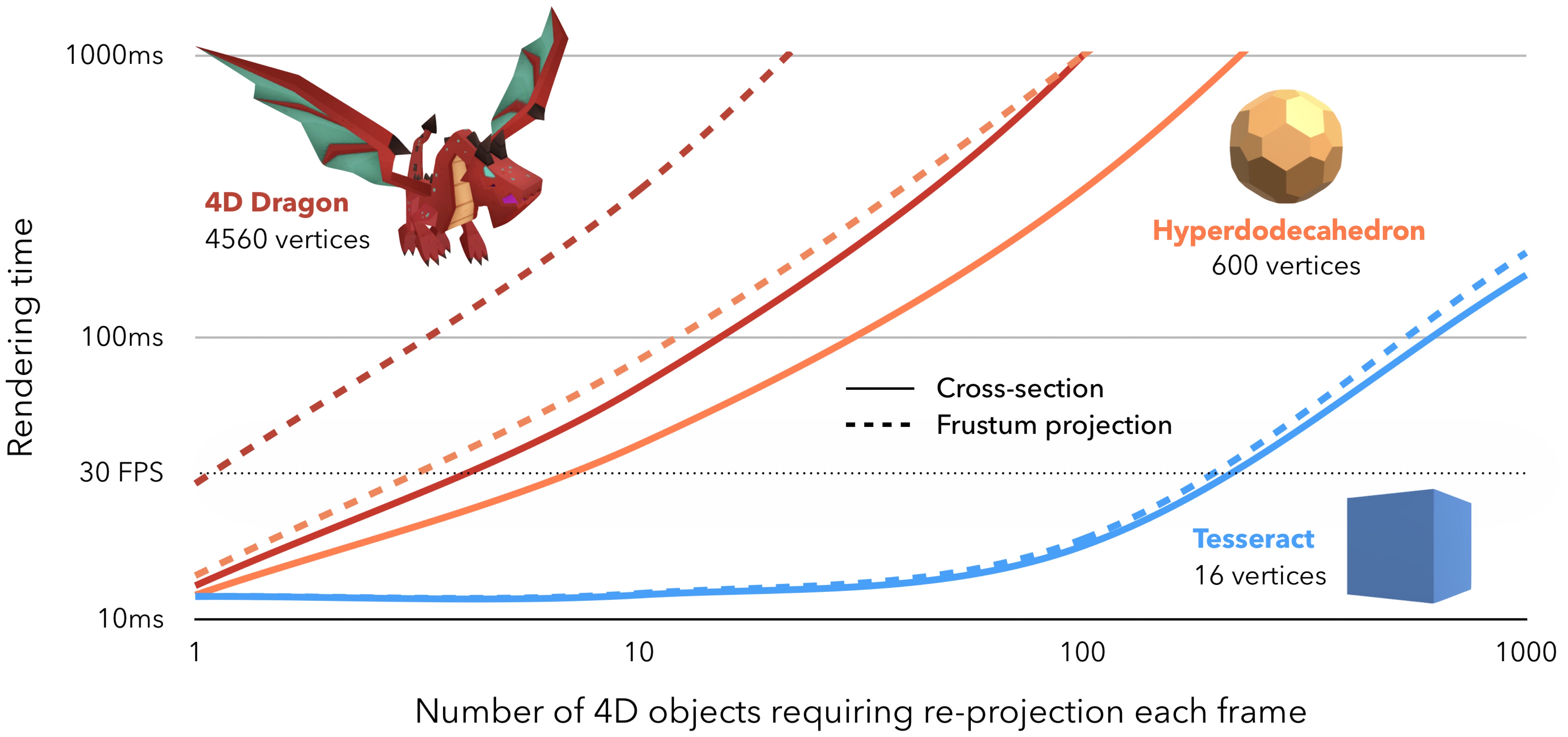}
  \vspace{-1.5em}
 \caption{\textbf{Rendering performance}. After initial projection, static 4D assets have the same performance footprint as 3D assets. However, animating such objects requires continuously recomputing their projected 3D meshes. The chart above shows how applying a continuous 4D rotation to different on-screen objects affects rendering performance in both cross-section and frustum projection. Performance tests have been measured on a 2018 MacBook Pro at full resolution.}
 \label{fig:performance}
 \vspace{-1em}
\end{figure}

\paragraph*{Performance Limitations.} 
Thanks to our \engine ~implementation, static 4D objects in the scene are treated as normal 3D assets after they are projected for the first time. However, when objects currently on-screen undergo a 4D transformation or the 4D camera is moved, the 3D projections of 4D objects need to be recomputed. As shown in Fig.~\ref{fig:performance}, mesh deformations take a big toll on overall rendering performance, especially when using frustum projection and 4D objects with higher numbers of vertices, and more sophisticated GPU implementations \cite{chu2009gl4d} should be considered. In particular, entering and exiting cross-section hyperplanes and rotating the 4D camera while using frustum projection are two of the most computationally expensive operations, since they require recomputing meshes dynamically. To circumvent these limitations and maintain a reasonable (>30 FPS) frame rate,  we designed \game ~as to prioritize the use of 4D primitives, and to make the user interact with no more than a couple complex 4D objects at a time.

\paragraph*{Parametric Explosion and Camera Control.}
While standard videogames and applications generally do not require more than a few forms of input to move the camera or interact with 3D objects, a 4D application can add up to twenty rototranslational variables atop already existing ones. This parametric explosion makes it particularly cumbersome to create fast-paced applications that allow full control of both 3D and 4D cameras and free manipulation of 4D objects. Therefore, we believe it makes sense to settle on allowing the user to control only a subset of these parameters, as we did by limiting the 4D camera movement to an orbital trajectory, or by constraining the axes for 4D manipulation. Instead of seeing this as a limitation, we believe choosing different subsets of parameters could lead to the creation of more unique applications with compelling mechanics.
In particular, it would be interesting to continue exploring the possible behaviors of the 4D camera, harnessing what is sometimes unpredictable visual potential. Due to the computationally expensive footprint, trade-offs such as alternating frustum projection with the much simpler cross-section should be explored. Even more importantly, dedicated visual aids should be provided to users, such as the 3D radar display and the 4D manipulation previews in \game.


\paragraph*{Making Higher Dimensions More Accessible.}
Despite science theories actually predict the existence of higher dimensions \cite{mcmullen2008visual}, conceiving of and visualizing higher dimensional entities has historically been a niche research topic, in which interest has rarely stemmed from potential usefulness---except for a few applications to crystallography and particle physics \cite{coxeter1973regular}.
Visualizing higher dimensions has been seen as an educational tool to better understanding our own 3D world, but 
its underlying technical complexity, level of abstraction, and fragmented resources make this topic very difficult to approach.
For this reason we aimed at presenting a different take on 4D visualization, as to \textit{make higher dimensions more accessible} to our community. Specifically, we tried to extend common 3D knowledge and present a self-contained paper with tangible examples, encompassing all major design considerations required to conceive a higher-dimensional world.
While our use case focuses on a videogame application, wherein game designers introduce unexpected scenarios and generate artistic content through 4D concepts, we hope the tools provided in this work can lead our readers to out-of-the-box thinking and inspire new creative applications to multiple domains.




%% file: conclusion.tex
\section{Conclusion}
In this work we built upon extant literature on higher dimensional visualization, taking a different path through the discipline: instead of focusing on the rendering and manipulation of 4D mathematical objects in a sandbox, we attempted to conceive of a much wider 4D universe---changing focus from the observation of a single geometry for educational purposes to a \textit{user-centric} exploration of a more complex high-dimensional narrative.
In particular, after observing the effects of reducing the dimensionality of 4D geometries, we studied the trade-offs related to defining a higher-dimensional universe in terms of geometrical structure, content generation, and authoring strategies---comparing the use of both cross-section and frustum projection approaches.
Finally, we showcased how multiple 4D and 3D objects can coexist and interact with each other in the same application, proposing a videogame as a concrete use case for how a four-dimensional engine might be integrated into existing rendering engines such as Unity3D.